\documentclass[showpacs,eqsecnum,aps]{revtex4}

\usepackage[dvips]{graphicx,color}

\begin{document}

\title{First Order Relativistic Three-Body Scattering}

\author{T. Lin, Ch. Elster}
\affiliation{
Institute of Nuclear and Particle Physics,  and
Department of Physics and Astronomy,  Ohio University, Athens, OH 45701}

\author{W. N. Polyzou}
\affiliation{
Department of Physics and Astronomy, The University of Iowa, Iowa City, IA
52242}

\author{ W. Gl\"ockle}
\affiliation{
Institute for Theoretical Physics II, Ruhr-University Bochum,
D-44780 Bochum, Germany}

\vspace{10mm}

\date{\today}

\begin{abstract}

Relativistic Faddeev equations for three-body scattering at arbitrary
energies are formulated in momentum space and in first order in the
two-body transition-operator directly solved in terms of momentum
vectors without employing a partial wave decomposition. Relativistic
invariance is incorporated within the framework of Poincar{\'e}
invariant quantum mechanics, and presented in some detail.  Based on a
Malfliet-Tjon type interaction, observables for elastic and break-up
scattering are calculated up to projectile energies of 1 GeV. The
influence of kinematic and dynamic relativistic effects on those
observables is systematically studied.  Approximations to the two-body
interaction embedded in the three-particle space are compared to the
exact treatment.

\end{abstract}

\vspace{10mm}

\pacs{21.45+v}

\maketitle





\section{Introduction}

Light nuclei can be accurately modeled as systems of nucleons interacting 
via effective two and three-body forces motivated e.g. by meson exchange.
This picture is expected to break down at a higher energy scale where
the physics is more efficiently described in terms of sub-nuclear
degrees of freedom.  One important question in nuclear physics is to
understand the limitations of models of nuclei as systems of
interacting nucleons.  Few-body methods have been an essential tool
for determining model Hamiltonians that describe low-energy nuclear
physics.  Few-body methods also provide a potentially useful framework
for testing the limitations of models of nuclei as few nucleon 
systems, however this requires extending the few-body models and
calculations to higher energy scales.  There are a number of
challenges that must be overcome to extend these calculations to
higher energies.  These include replacing the non-relativistic theory
by a relativistic theory, limitations imposed by interactions fit to
elastic scattering data, new degrees of freedom that appear above the
pion production threshold, as well as numerical problems related to
the proliferation of partial wave at high energies.  In this paper we
address some of these questions.  We demonstrate that it is possible
to now perform relativistic three-body scattering calculations at
energies up to 1 GeV laboratory kinetic energy.  The key elements of
our success is the use of direct integration methods that avoid the use
of partial waves and new techniques for treating functions of non-commuting 
operators that appear in the relativistic nucleon-nucleon interactions.

During the last two decades calculations of nucleon-deuteron
scattering based on momentum-space Faddeev equations \cite{faddeev}
experienced large improvements and refinements.  It is fair to state
that below about 200~MeV projectile energy the momentum-space Faddeev
equations for three-nucleon (3N) scattering can now be solved with
high accuracy for realistic two- and three-nucleon interactions. A
summary of these achievements can be found in
Refs.~\cite{wgphysrep,wgarticle,kuros,sauer}. The approach described
there is based on using angular momentum eigenstates for the two- and
three-body systems. This partial wave decomposition replaces the
continuous angle variables by discrete orbital angular momentum
quantum numbers, and thus reduces the number of continuous variables
to be discretized in a numerical treatment.  For low projectile
energies the procedure of considering orbital angular momentum
components appears physically justified due to arguments related to
the centrifugal barrier and the short range of the nuclear force.  If
one considers three-nucleon scattering at a few hundred MeV projectile
kinetic energy, the number of partial waves needed to achieve convergence
proliferates, and limitations with respect to computational
feasibility and accuracy are reached.  It appears therefore natural to
avoid a partial wave representation completely and work directly with
vector variables. This is common practice in bound state calculations
of few-nucleon systems based on variational \cite{Arriaga95a} and
Green's function Monte Carlo (GFMC) methods
\cite{Carlson87a,Carlson88a,Zabolitzki82a,Carlson99a} and was for the
first time applied in momentum space Faddeev calculations 
for  bound states in \cite{bound3d} and for 
scattering at intermediate energies in Ref.~\cite{scatter3d}.

The key advantage of a formulation of the Faddeev equations in terms
of vector variables lies in its applicability at higher energies,
where special relativity is expected to become relevant. Poincar{\'e} 
invariance is an exact symmetry that should be
satisfied by all calculations, however in practice consistent relativistic
calculations are more numerically intensive, thus making their
nonrelativistic counterpart a preferred choice. Furthermore, estimates
of relativistic effects have been quantitatively small for 3N
scattering below 200 MeV \cite{witalar1,witalar2,witalar3} with the
exception of some breakup cross sections in certain phase space
regions~\cite{Skibin}, indicating that at those energies
non-relativistic calculations have sufficient precision.  This is in
part because in either a relativistic or non-relativistic model the
interactions are designed to fit the same invariant differential cross
section \cite{moller}, which can be evaluated in any frame using
standard kinematic Lorentz transformations, so model calculations are
designed so that there are no ``relativistic corrections'' at the
two-body level.  Three-body interactions can even be chosen so the
non-relativistic calculations fit both the two- and three-body
invariant cross sections.  This can be done in one frame and the
invariance of the cross section fixes it in all other frames using
standard relativistic kinematics.  This, procedure has internal
inconsistencies which show up if these models are used 
as input in larger systems, but they clearly indicate that the
problem of identifying relativistic effects is more subtle than simply
computing non-relativistic limits.  In this paper we focus on
differences between relativistic and non-relativistic calculations
with two-body input that have the same cross section and use the same
two-body wave functions \cite{KG,cps,keipo}.

There are two primary approaches for modeling relativistic few-body
problems.  One treats Poincar\'e invariance as a symmetry of a quantum
theory, the other is based on quasipotential reductions
\cite{Frohlich:1983} of formal relations \cite{Dyson:1949,
Schwinger:1951} between covariant amplitudes. One specific realization
of this approach is the covariant spectator approach of
Ref.~\cite{stadler}.  In this paper the relativistic three-body
problem is formulated within the framework of Poincar{\'e} invariant
quantum mechanics.  It has the advantage that the framework is valid
for any number of particles and the dynamical equations have the same
number of variables as the corresponding non-relativistic equations.
Poincar\'e invariance is an exact symmetry that is realized by a
unitary representation of the Poincar\'e group on a three-particle
Hilbert space.  The dynamics is generated by a Hamiltonian.  This
feature is shared with the Galilean invariant formulation of
non-relativistic quantum mechanics.  The Hamiltonian of the
corresponding relativistic model differs in how the two-body
interactions are embedded in the three-body center of momentum
Hamiltonian (mass operator).  The equations we use to describe the
relativistic few-problem have the same operator form as the
nonrelativistic ones, however the ingredients are different.

In this article we want to concentrate on the leading order term of
the Faddeev multiple scattering series within the framework of
Poincar{\'e} invariant quantum mechanics. The first order term
contains already most relativistic ingredients which, together with
the relativistic free three-body resolvent, gives the kernel of the
integral equation. We want to understand essential differences between
a relativistic and nonrelativistic approach already on the basis of
the first order term.  As simplification we consider three-body
scattering with spin-independent interactions.  This is mathematically
equivalent to three-boson scattering.  The interactions employed are
of Yukawa type, and no separable expansions are employed.  In order to
obtain a valid estimate of the size of relativistic effects, it is
important that the interactions employed in the nonrelativistic and
relativistic calculations are phase-shift equivalent. To achieve this
we employ here the approach suggested by Kamada-Gl\"ockle \cite{KG},
which uses a unitary rescaling of the momentum variables to change the
nonrelativistic kinetic energy into the relativistic kinetic energy.

This article is organized as follows.  Section II discusses the
formulation of Poincar\'e invariant quantum mechanics, and section III
discusses the structure of the dynamical two- and three-body mass
operators.  Scattering theory formulated in terms of mass operators is
discussed in section IV.  The formulation of the Faddeev equations and
techniques for computing the Faddeev kernel are discussed in section V.
Details on kinematical aspects of how to construct the cross sections
is given in section VI.  In Sections VIII and IX we present
calculations for elastic and breakup processes in the intermediate
energy regime from 0.2 to 1 GeV. Our focus here is to compare
different approximations to the embedded interaction with respect to
the exact calculation.  Our conclusions are in Section X.  Two
Appendices are devoted to relating the transition matrix elements based
on mass operators to the invariant amplitudes with the conventions used in
the particle data book and expressing the invariant cross section and
differential cross sections worked out directly in laboratory-frame variables.


\section{Poincar\'e Invariant Quantum Mechanics}

Symmetry under change of inertial coordinate system is the fundamental
symmetry of Poincar\'e invariant quantum mechanics.  In special relativity 
different inertial coordinate systems are related by the subgroup of Poincar\'e
transformations continuously connected to the identity.  In this paper
the Poincar\'e group refers to this subgroup, which excludes the
discrete transformations of space reflection and time reversal.
Wigner \cite{wigner} proved that a necessary and sufficient condition
for quantum probabilities to be invariant under change of inertial
coordinate system is the existence of a unitary representation, ${\cal
U}(\Lambda ,a)$, of the Poincar\'e group on the model Hilbert space.
Equivalent vectors, $\vert \psi \rangle$ and $\vert \psi' \rangle$, in
different inertial coordinate systems are related by:
\begin{equation}
\vert \psi' \rangle = {\cal U}(\Lambda ,a) \vert \psi \rangle .
\label{eq:b.1}
\end{equation}

In Poincar\'e invariant quantum mechanics the dynamics is generated by
the time evolution subgroup of ${\cal U}(\Lambda ,a)$.  The
fundamental dynamical problem is to decompose ${\cal U}(\Lambda ,a)$
into a direct integral of irreducible representations.  This is the
analog of diagonalizing the Hamiltonian or time-evolution operator in
non-relativistic quantum mechanics.  The problem of formulating the
dynamics is to construct the dynamical representation ${\cal
U}(\Lambda ,a)$ of the Poincar\'e group by introducing realistic
interactions in the tensor products of single particle irreducible
representations in a manner that preserves the group representation
property and essential aspects of cluster separability.  The solution
to this non-linear problem is achieved by adding suitable interactions
to the Casimir operators of non-interacting irreducible representations
of the Poincar\'e group.

Since irreducible representations of the Poincar\'e group play a
central role in both the formulation and solution of the dynamical
model, we give a brief summary of the construction of the irreducible
representations that we use in this paper.  The Poincar\'e group is a
ten parameter group that is the semidirect product of the Lorentz
group and the group of spacetime translations.  Spacetime translations
are generated by the four momentum operator, $P^{\mu}$, and Lorentz
transformations are generated by the antisymmetric angular momentum
tensor, $J^{\mu \nu}$.   

The Pauli Lubanski vector is the four vector operator defined by 
\begin{equation} 
W^{\mu} = -{1 \over 2} \epsilon^{\mu\alpha \beta \gamma}
P_\alpha J_{\beta \gamma}. 
\label{eq:b.2}
\end{equation} 
The Casimir operators for the Poincar\'e group are 
\begin{equation}
M^2 = -\eta_{\mu \nu} P^{\mu}P^{\nu}=H^2 - \mathbf{P} \cdot \mathbf{P}
\label{eq:b.3}
\end{equation} 
and 
\begin{equation}  
W^2 = -\eta_{\mu \nu} W^{\mu}W^{\nu} = M^2 j^2
\label{eq:b.4}
\end{equation}
where $\eta_{\mu\nu}$ is the Minkowski metric, $M$ is the mass operator,
$H$ is the Hamiltonian, $\mathbf{P}$ is the linear momentum, and
$j^2$ is the spin. 

Positive-mass positive-energy irreducible representations are labeled
by eigenvalues of the mass $M$ and spin $j^2$.  Vectors in an
irreducible subspace are square integrable functions of the
eigenvalues of a complete set of commuting Hermitian operator-valued
functions of the generators $P^{\mu}$ and $J^{\mu\nu}$.  In addition
to the two invariant Casimir operators, it is possible to find four
additional commuting Hermitian functions of the generators.  For each
of these four commuting observables it is possible to find conjugate
operators.  These conjugate operators, along with the eigenvalues of
the Casimir operators, fix the eigenvalue spectrum of the four
commuting Hermitian operators.  The irreducible representation space
is the space of square integrable functions of the eigenvalues of the
four commuting operators.  The generators can be expressed as
functions of these four operators, their conjugates, and the Casimir
invariants \cite{fcwp,bkwp,wp2}.

In this paper we choose the four commuting operators to be the three
components of the linear momentum and the $z$ component of the
canonical spin operator.  In this representation the four conjugate
operators are taken as the partial derivatives of the momentum holding
the canonical spin constant (Newton-Wigner position \cite{nw}
operator) and the $x$ component of the canonical spin.  While $j_x$ is
not exactly conjugate to $j_z$, the two operators generate the full
$SU(2)$ spin algebra.  The corresponding eigenstates have the form
\begin{equation}
\vert \mathbf{p}, \mu \rangle :=\vert (m,j) \mathbf{p}, \mu \rangle .
\label{eq:b.5}
\end{equation} 

The mass $m$ spin $j$ irreducible representation of the Poincar\'e
group in this basis is determined from the group
representation property and the action of rotations, spacetime
translations and canonical boosts on the zero momentum eigenstates:

\begin{equation} 
{\cal U}(R,0) \vert \mathbf{0}, \mu \rangle =
\vert  \mathbf{0}, \mu' \rangle D^j_{\mu'\mu} (R) 
\label{eq:b.6}
\end{equation}
\begin{equation}
{\cal U}(I,a) \vert  \mathbf{0}, \mu \rangle =
e^{-i a^0 m} \vert  \mathbf{0}, \mu \rangle
\label{eq:b.7}
\end{equation}
\begin{equation}
{\cal U}(B(p_m) ,0) \vert  \mathbf{0}, \mu \rangle =
\vert \mathbf{p}, \mu \rangle \sqrt{{E_{p_m}\over m}} 
\label{eq:b.8}
\end{equation}
where in these equations $R$ is a rotation, $D^j_{\mu'\mu} (R)$ is the
standard $2j+1$ dimensional unitary representation of $SU(2)$,
$a=(a^0,\mathbf{a})$ is a displacement four vector, $B(p_m)$
is the rotationless Lorentz boost (canonical boost) that transforms 
$(m,\mathbf{0})$ to
$p_m := (E_{p_m},\mathbf{p})$, 
\begin{equation}
(B (p_m))^{\mu}{}_{\nu}  :=
\left (
\begin{array}{cc}
{E_{p_m} / m}  & {\mathbf{p} / m}\\
{\mathbf{p} / m} & I + {\mathbf{p}\otimes
\mathbf{p} \over m(m+E_{p_m})}
\end{array}
\right ),
\label{eq:b.9}
\end{equation}
and $E_{p_m} = \sqrt{m^2 + \mathbf{p}^2}$.  That the magnetic quantum
number remains invariant in (\ref{eq:b.8}) under the rotationless
boost (\ref{eq:b.9}) is the defining
property of the canonical spin.

The multiplicative factor on the right side of (\ref{eq:b.8}) is 
fixed up to a phase by unitarity and the normalization convention
\begin{equation}
\langle \mathbf{p}\,' ,\mu' \vert \mathbf{p} , \mu \rangle
= \delta (\mathbf{p}\,' - \mathbf{p}) \delta_{\mu' \mu} .
\label{eq:b.11}
\end{equation}

With these choices the action of an arbitrary Poincar\'e transformation 
on these states is given by 
\begin{equation}
{\cal U}_{mj}(\Lambda ,a ) \vert \mathbf{p} , \mu \rangle
= \vert \mathbf{p}\,' , \mu' \rangle
\sqrt{E_{p_m}' \over
E_{p_m} } D^j_{\mu' \mu}\left (R_w (\Lambda ,p_m)\right )
e^{i p_m' \cdot a} ,
\label{eq:b.12}
\end{equation}
where 
$R_w (\Lambda ,p_m)$ is the (standard) Wigner
rotation,
\begin{equation}
R_w (\Lambda ,p_m) := B^{-1} (\Lambda p_m)\Lambda B(p_m),
\label{eq:b.13}
\end{equation}
and $p_m'=\Lambda p_m$.  Since each of the elementary transformations 
(\ref{eq:b.6},\ref{eq:b.7},\ref{eq:b.8}) are unitary, it follows that 
(\ref{eq:b.12}) is unitary.  Since every basis vector can be generated
from the $\mu=j$ and $\mathbf{P}=\mathbf{0}$ basis vector using equations 
(\ref{eq:b.6},\ref{eq:b.7},\ref{eq:b.8}), representation
(\ref{eq:b.12}) is also irreducible.

The mass $m$ spin $j$ irreducible representations that are used in
this paper have the form (\ref{eq:b.12}).  The mass $m$ spin $j$
irreducible representation space with basis (\ref{eq:b.5}) is denoted
by ${\cal H}_{mj}$.

The Hilbert space for the-three nucleon problem is the tensor product
of three one-nucelon irreducible representation spaces:
\begin{equation}
{\cal H} ={\cal H}_{m{1\over 2}} \otimes {\cal H}_{m{1\over 2}} 
\otimes {\cal H}_{m{1\over 2}} .
\label{eq:b.14}
\end{equation}
In this paper all nucleons are assumed to have the same mass, $m$. 

The non-interacting unitary representation of the Poincar\'e group
on ${\cal H}$ is the tensor product of three one-nucleon irreducible
representations:
\begin{equation}
{\cal U}_0 (\Lambda ,a) = {\cal U}_{m{1\over 2}} (\Lambda ,a) \otimes
{\cal U}_{m{1\over 2}} (\Lambda ,a) \otimes
{\cal U}_{m{1\over 2}} (\Lambda ,a) .
\label{eq:b.15}
\end{equation}

As in the case of rotations, the tensor product of irreducible
representations of the Poincar\'e group is reducible.  The tensor
product of three irreducible representations can be decomposed into a
direct integral of irreducible representation using Clebsch-Gordan
coefficients for the Poincar\'e group.  The Clebsch-Gordan
coefficients for the Poincar\'e group are known \cite{stora,fch,bkwp}.
As in the case of rotations, the Poincar\'e Clebsch-Gordan
coefficients are basis dependent and the three-body irreducible
representations can be generated by pairwise coupling.  The Poincar\'e
Clebsch-Gordan coefficients can be computed by using $SU(2)$
Clebsch-Gordan coefficients to decompose three-body zero momentum
eigenstates states into irreducible $SU(2)$ representations.
Three-body irreducible basis vectors are generated by applying 
Eqs. (\ref{eq:b.6},\ref{eq:b.7},\ref{eq:b.8}) to the zero-momentum
$SU(2)$ irreducible representations.

The resulting irreducible three-body basis depends on the order of the
coupling.  In the basis of eigenstates of the three-body linear
momentum and canonical spin the irreducible eigenstates are labeled by
eigenvalues ${\sf W}$ of the three-body invariant mass, $M_0$, the
three-body canonical spin, $j$ (for simplicity of notation we use the
same label $j$ for the total canonical spin of the three-body system
and the single particle canonical spin), the total three-body
momentum, $\mathbf{P}$, the $z$ component of the three-body canonical
spin, $\mu$, and invariant degeneracy quantum numbers, $d$, that
distinguish multiple copies of the same irreducible representation:
\begin{equation}
\vert ({\sf W},j), \mathbf{P} , \mu , d \rangle .
\label{eq:b.16}
\end{equation}
For two-particle systems the degeneracy quantum numbers $d$ are
discrete (for example they may be taken to be invariant spin and orbital
angular momentum quantum numbers) while for more than two particles
the degeneracy quantum numbers will normally include invariant
sub-energies, which have a continuous eigenvalue spectrum.
In addition to the appearance of the degeneracy quantum numbers, the 
eigenvalue spectrum of the free invariant mass operator, $M_0$, 
is continuous.

These states transform as mass ${\sf W}$ spin $j$ irreducible
representations of the Poincar\'e group under ${\cal U}_0(\Lambda
,a)$:
\begin{equation}
{\cal U}_0(\Lambda ,a ) \vert ({\sf W},j), \mathbf{P} , \mu , d \rangle
= \vert ({\sf W},j), \mathbf{P}\,' , \mu' , d \rangle
\sqrt{ E_{P_{\sf W}}' \over E_{P_{\sf W}} }
D^j_{\mu' \mu}\left (R_w (\Lambda ,P_{\sf W})\right )
e^{i P_{\sf W}' \cdot a} ,
\label{eq:b.17}
\end{equation}
where
\begin{equation}
P'_{\sf W} = \Lambda P_{\sf W}
\qquad P_{\sf W}:= (\sqrt{{\sf W}^2 + \mathbf{P}^2 },\mathbf{P})=
(E_{P_{\sf W}}, \mathbf{P}).
\label{eq:b.18}
\end{equation}
The quantities ${\sf W},j,d$ are invariants of the representation
(\ref{eq:b.17}) of ${\cal U}_0(\Lambda ,a)$.

Because the Poincar\'e group allows time evolution to be expressed in
terms of spatial translations and Lorentz boosts, when particles
interact, consistency of the initial value problem requires that the
unitary representation of the Poincar\'e group depends non-trivially
on the interactions.  The construction of ${\cal U} (\Lambda ,a)$ for
dynamical models is motivated by the example of Galilean invariant
quantum mechanics.  The non-relativistic three-body Hamiltonian has
the form
\begin{equation}
H = {\mathbf{P}^2 \over 2M_g} + h
\label{eq:b.19}
\end{equation}
where the Casimir Hamiltonian, $h$, is the Galilean invariant part of the
Hamiltonian and $M_g$ is the Galilean mass.  In the non-relativistic
case interactions are added to the non-interacting Casimir Hamiltonian
$h_0$ :
\begin{equation}
h=h_0+V_{nr}
\label{eq:b.20}
\end{equation}
where the Galilean invariance of $h$ requires that 
the interaction $V_{nr}$ is rotationally invariant and commutes
with and is independent of the linear momentum $\mathbf{P}$.  This
means that in the corresponding non-relativistic basis
\begin{equation}
\langle {\sf h}, \mathbf{P} , j ,\mu ,d \vert V_{nr} \vert
{\sf h}', \mathbf{P}\,' , j' ,\mu' ,d' \rangle
=
\delta( \mathbf{P}- \mathbf{P}\,') \delta_{jj'} \delta_{\mu \mu'}
\langle {\sf h},d \Vert V^j_{nr} \Vert
{\sf h}' ,d' \rangle .
\label{eq:b.21}
\end{equation}
where ${\sf h}$ is an eigenvalue of $h_0$.

In the Poincar\'e invariant case Eq.~(\ref{eq:b.19}) is replaced by
\begin{equation}
H = \sqrt{\mathbf{P}^2 + M^2}
\label{eq:b.22}
\end{equation}
where $M$ in Eq.~(\ref{eq:b.22}) plays the same role as the Casimir
Hamiltonian $h$ in Eq.~(\ref{eq:b.19}). The corresponding free
relativistic Hamiltonian is $H_0=\sqrt{\mathbf{P}^2 + M^2_0}$.
In what follows $\lambda$ denotes the eigenvalue of $M$
to distinguish it from the eigenvalue
${\sf W}$ of $M_0$.

A Poincar\'e invariant dynamics can be constructed by adding an
interaction to the non-interacting $M_0$ that commutes with and is
independent of $\mathbf{P}$ and $j_z$:
\begin{equation}
M= M_0 + V_r .
\label{eq:b.23}
\end{equation}
In the non-interacting irreducible basis (\ref{eq:b.16}) these
conditions require interactions of the form
\begin{equation}
\langle ({\sf W},j), \mathbf{P},\mu ,d \vert V_r \vert
({\sf W}',j'), \mathbf{P}\,' ,\mu',d' \rangle
=
\delta( \mathbf{P}- \mathbf{P}\,') \delta_{jj'} \delta_{\mu \mu'}
\langle {\sf W},d \Vert V^j_{r} \Vert
{\sf W}' ,d' \rangle .
\label{eq:b.24}
\end{equation}

The dynamical problem is to find simultaneous eigenstates of the
commuting operators $M$, $\mathbf{P}$, $j_z$, and $j^2$.  This is
done by diagonalizing $M$ in the irreducible basis (\ref{eq:b.16}).
The eigenfunctions of $M$ have the form
\begin{equation}
\langle ({\sf W},j), \mathbf{P} , \mu , d  \vert
(\lambda ,j'), \mathbf{P}\,' , \mu' \rangle =
\delta(\mathbf{P}-\mathbf{P}\,')
\delta_{jj'} \delta_{\mu \mu'}  
\langle {\sf W},d ,j  \vert \lambda ,j \rangle 
\label{eq:b.25}
\end{equation}
where the eigenfunctions $\langle {\sf W},d ,j  \vert \lambda ,j \rangle$
are solutions of 
\begin{equation}
{\sf W} \langle {\sf W},j,d \vert \lambda ,j \rangle
+ \sum'  \int'
\langle {\sf W},d \Vert V^j_{r} \Vert
{\sf W}' ,d' \rangle d{\sf W}' dd'
\langle {\sf W}',j,d' \vert \lambda ,j \rangle =
\lambda \langle {\sf W},j,d \vert \lambda ,j \rangle ,
\label{eq:b.26}
\end{equation}
with mass eigenvalue $\lambda$.  For two-particle systems the
degeneracy quantum numbers $d$ are discrete (for example they may
taken to be invariant spin and orbital angular momentum quantum
numbers) while for more than two particles the degeneracy quantum
numbers will normally include invariant sub-energies, which have a
continuous eigenvalue spectrum.

Because $\{ M_0,j^2,j_z,\mathbf{P},j_{x},-i\mbox{\boldmath$\nabla$}_P
\} $ have the same commutations relations as $\{
M,j^2,j_z,\mathbf{P},j_{x},-i \mbox{\boldmath$\nabla$}_P \}$, if the
dynamical Poincar\'e generators are defined as the same functions of
these operators \cite{wp2,bkwp} with $M_0$ replaced by $M$, it
follows that the simultaneous eigenstates
\begin{equation}
\vert (\lambda ,j), \mathbf{P} , \mu \rangle
\label{eq:b.27}
\end{equation}
of $ M,j^2,\mathbf{P},j_z$ transform as a mass $\lambda$ spin $j$
irreducible representation of the Poincar\'e group.  

Since these eigenstates are complete, this defines the dynamical
representation of the Poincar\'e group on a basis by
\begin{equation}
{\cal U} (\Lambda ,a ) \vert (\lambda,j), \mathbf{P} , \mu  \rangle
= \vert (\lambda,j), \mathbf{P}\,' , \mu'  \rangle
\sqrt{E_{P_\lambda}' \over E_{P_\lambda} } D^j_{\mu' \mu}\left
(R_w (\Lambda ,P_{\lambda})\right )
e^{i P'_{\lambda} \cdot a}  ,
\label{eq:b.28}
\end{equation}
where
\begin{equation}
E_{P_\lambda} =
\sqrt{\lambda^2 + {\mathbf P}^2}, \qquad P_{\lambda}:= (E_P,{\mathbf P}),
\qquad P'_{\lambda} = \Lambda P_{\lambda}.
\label{eq:b.29}
\end{equation}
This shows that the solution to the eigenvalue problem (\ref{eq:b.26})
provides the desired decomposition of the dynamical unitary representation of
the Poincar\'e group into a direct integral of irreducible
representations.

The appearance of the mass eigenvalue $\lambda$ on the right side of
Eq.~(\ref{eq:b.28}) indicates the interaction dependence of this
representation.  It can happen, for a given choice of irreducible
basis, that the coefficient on the right hand side of (\ref{eq:b.28})
is independent of the mass eigenvalue for a subgroup of the Poincar\'e
group.  For the basis (\ref{eq:b.16}), of linear momentum and canonical
spin eigenstates, both translations and rotations have this property.
These transformations generate a three-dimensional Euclidean subgroup
of the Poincar\'e group that is independent of the interaction.  An
interaction-independent subgroup is called a kinematic subgroup; the
three-dimensional Euclidean group is the kinematic subgroup is 
associated with an instant-form dynamics \cite{dirac}.

\section{Mass operators} 

Next we discuss the structure of mass operators for the two
and three-body problems.  We pay particular attention to issues 
related to representations of these operators that are 
suitable for computations without using partial waves.

The construction of the dynamics in (\ref{eq:b.26}) and
(\ref{eq:b.28}) adds an interaction to the mass Casimir operator of a
non-interacting irreducible representation of the Poincar\'e group to
construct an interacting irreducible representation.  The role of the
spin in the structure of the irreducible representations suggests that
this construction requires a partial wave decomposition, however
partial waves are not used in the calculations that follow.

The spin in the relativistic case is obtained by coupling the single
particle spins and orbital angular momentum vectors.  While the form
of the coupling is more complex than it is in the non-relativistic
case, the final step involves coupling redefined spins and orbital
angular momenta with ordinary $SU(2)$ Clebsch-Gordon coefficients.
Undoing this coupling leads to a representation of
the dynamical operators that can be used in a calculation based on
vector variables.  

The first step is to construct redefined vector variables that can be
coupled to obtain the spin.  To understand the transformation 
properties of these
operators note that the magnetic quantum number in equation
(\ref{eq:b.28}) undergoes a Wigner rotation when the system is Lorentz
transformed.  If the spin of the representation is obtained by
coupling the redefined particle spins and orbital angular momenta with
SU(2) Clebsch-Gordan coefficients, then all of the spins and relative
angular momenta must also transform with the same Wigner rotation.

To illustrate how to construct momentum operators that
Wigner rotate under kinematic Lorentz boosts consider a pair of noninteracting
spinless particles.  The total four momentum $P_{M_0}$ of this system is the
sum of the single particle four momenta
\begin{equation}
P_{M_0} = p_{m_1} + p_{m_2}.
\end{equation}
Define the operator $k$ by
\begin{equation}
k:= B^{-1} (P_{M_{0}}) p_{m_1} ,
\label{eq:b.30}
\end{equation}
where
$B^{-1} (P_{M_{0}})$ is interpreted as a $4\times 4$ 
matrix of {\it operators}.
If both $P_{M_{0}}$ and $p_{m_1}$ are transformed with a Lorentz
transformation $\Lambda$, then $k$ rotates with the same 
Wigner rotation that appears in Eq.~(\ref{eq:b.17})
\begin{equation}
k' = B^{-1} (P_{M_{0}}') p_1' =
B^{-1} (\Lambda P_{M_{0}}) \; \Lambda p_1 =
B^{-1} (\Lambda P_{M_{0}}) \Lambda B(P_{M_{0}})
B^{-1} (P_{M_{0}}) p_1 =
R_w (\Lambda ,P_{M_{0}}) \; k .
\label{eq:b.31}
\end{equation}
It is due to the operator nature of $B^{-1}(P_{M_0})$ that
$k$ does not transform as a 4-vector.

The tensor product of single particle basis vectors 
$\vert \mathbf{p}_1, \mathbf{p}_2 \rangle$
can be replaced by a basis $\vert \mathbf{P}, \mathbf{k} \rangle$
using a variable change.  In this basis Eq.~(\ref{eq:b.31}) 
implies
\begin{equation}
{\cal U}_0 (\Lambda,a)
\vert \mathbf{P}, \mathbf{k} \rangle =
\vert \mathbf{P}', R_w (\Lambda ,P_{\sf W}) \mathbf{k} \rangle
\sqrt{E_{P_{\sf W}}' \over E_{P_{\sf W} }}  e^{i P_{\sf W}' \cdot a}
\qquad P'_{\sf W}=\Lambda
P_{\sf W}
\label{eq:b.32}.
\end{equation}
where $\mathbf{k}$ is the eigenvector of the space components of the 
operator (\ref{eq:b.30}).  This shows that $\mathbf{k}$ undergoes the 
same Wigner-rotation as the two-body canonical spin. 

If the two particles have spin, this single particle spins need to be
Wigner rotated before they can be coupled \cite{bkwp}.  The spins
obtained this way are called constituent spins. The constituent spins
undergo the same Winger rotations as $\mathbf{k}$ but they are not
1-body operators and do not have natural couplings to the
electromagnetic interaction.
In this paper we only consider spinless interactions.  In this case
the constituent spins can be ignored.  Their only effect is that they
impact the permutation symmetry of orbital-isospin part of the wave
functions.  

The magnitude of $\mathbf{k}$ is an invariant that fixes the
two-body invariant mass eigenvalue ${\sf W}$:
\begin{equation}
W= 2 \sqrt{\mathbf{k}^2 + m^2} .
\label{eq:b.33}
\end{equation}
If the vector part of
$\mathbf{k}$ is expanded in partial waves
\begin{equation}
\vert \mathbf{P}, \vert \mathbf{k} \vert ,j, \mu \rangle :=
\int \vert \mathbf{P} ,\mathbf{k} \rangle \,
Y^{j\mu} (\hat{\mathbf{k}}) \, d \hat{\mathbf{k}}
\label{eq:b.34}
\end{equation}
then
\begin{equation}
{\cal U}_0 (\Lambda,a)\vert \mathbf{P}, \vert \mathbf{k} \vert, j, \mu
\rangle := \sum_{\mu'}  \vert \mathbf{P}\,', \vert
\mathbf{k} \vert, j, \mu' \rangle \sqrt{E_{P_{\sf W}}' \over E_{P_{\sf W}} }
D^j_{\mu'\mu}(R_w (\Lambda ,P_{\sf W})) e^{i P_{\sf W}' \cdot a}
\label{eq:b.35}
\end{equation}
transforms like (\ref{eq:b.12}). 

In the representation
$\vert \mathbf{P}, \mathbf{k} \rangle$
Eq.~(\ref{eq:b.24}) is satisfied for 
a spinless interaction of the form
\begin{equation}
\langle \mathbf{P}, \mathbf{k}  \vert V_r \vert
\mathbf{P}\,'  , \mathbf{k}' \rangle
=
\delta( \mathbf{P}- \mathbf{P}\,')
\langle \mathbf{k} \Vert V_{r} \Vert
\mathbf{k}' \rangle
\label{eq:b.36}
\end{equation}
with a rotationally invariant kernel 
\begin{equation}
\langle R\mathbf{k} \Vert V_{r} \Vert
R \mathbf{k}' \rangle = \langle \mathbf{k} \Vert V_{r} \Vert
\mathbf{k}' \rangle .
\end{equation} 
If the interaction includes nucleon spins, the rotational invariance 
must be generalized to include rotationally invariant contributions 
involving the constituent spins.

Next we consider the three-body problem, where
${\cal U}_0(\Lambda,a)$, $M_0$ and ${\sf W}$ are now associated
with the three nucleon system.
In the three-body system vector operators that Wigner rotate are
the Poincar\'e-Jacobi momenta and three-body constituent spins.  
The Jacobi momenta are obtained from the
non-relativistic Jacobi momenta by replacing Galilean boosts to the
zero momentum frame of a system or subsystem by the corresponding
non-interacting Lorentz boosts.   In these expression the boosts
are considered to be matrices of operators.
The replacements are
\begin{eqnarray}
\tilde{\mathbf{k}}_{ij} \equiv 
B_g^{-1} (\mathbf{P}) (\mathbf{p}_i - \mathbf{p}_j) =
 \mathbf{p}_i - \mathbf{p}_j
 - \frac{({\bf p}_i+{\bf p}_j)}{(m_i+m_j)} (m_i-m_j)
&\longrightarrow &
\tilde{\mathbf{k}}_{ij} \equiv B^{-1} (P_{m_{0,ij}} )( p_i-p_j)
\label{eq:b.37} \\
\mathbf{q}_i \equiv  
B_g^{-1} (\mathbf{P}) \mathbf{p}_i = 
\mathbf{p}_i - {\mathbf{P} \over M_g} m_i
& \longrightarrow &
\mathbf{q}_i \equiv B^{-1} (P_{M_0} ) p_i .
\label{eq:b.38}
\end{eqnarray}
where $m_{0,ij}$ is the invariant mass of the noninteracting 
two-particle $(i,j)$ system.

The only complication in the three-body case is that when the
single particle momenta undergo Lorentz transformations
the variables $\tilde{\mathbf{k}}_{ij}$, $\mathbf{q}_i$ experience
{\it different} Wigner rotations
\begin{eqnarray}
\mathbf{q}_i &\to& R_w(\Lambda , P_{M_0}) \mathbf{q}_i
\label{eq:b.39}\\
\tilde{\mathbf{k}}_{ij} &\to& R_w(\Lambda , P_{m_{0,ij}} )
\tilde{\mathbf{k}}_{ij}.
\label{eq:b.33b}
\end{eqnarray}
Because of the different Wigner rotations, the angular momenta
associated with $\mathbf{q}_i$ and $\tilde{\mathbf{k}}_{ij}$
cannot be consistently 
coupled with $SU(2)$ Clebsch-Gordon coefficients.  To fix this {\it
redefine} $\tilde{\mathbf{k}}_{ij} \to \mathbf{k}_{ij}$ by replacing
all of the ${p}_i$s in (\ref{eq:b.37}) by the corresponding ${q}_i$s.
Then when the single particle variables are Lorentz transformed, the
$\mathbf{q}_i$ all Wigner rotate with a rotation $R'$.  This means
that the redefined $\mathbf{k}_{ij}$ transform as $R_w ( R',
q_{ij})$, where $q_{ij} = q_i + q_j$.  
But the {\it defining property} of the canonical boost
(\ref{eq:b.9}) is $R_w ( R', q_{ij})= R'$ which means that both
$\mathbf{q}_{i}$ and $\mathbf{k}_{ij}$ undergo the same Wigner
rotation, $R'$, when the single-particle variables are Lorentz
transformed.

Only two of the six vector variables, $\mathbf{q}_{i}$ and
$\mathbf{k}_{ij}$, are linearly independent.  Any two of these
variables along with $\mathbf{P}$ can be used to label three-body
basis vectors.  Following  non-relativistic usage, the single
particle momenta are replaced by the independent variables
\begin{equation}
\{ \mathbf{P} , \mathbf{q}_{k}, \mathbf{k}_{ij} \}
\label{eq:b.40}
\end{equation}
where $k \not=i,j$.  The single particle basis vectors are 
replaced by
\begin{equation}
\vert \mathbf{P} , \mathbf{q}_k , \mathbf{k}_{ij} \rangle =
\vert \mathbf{p}_1 , \mathbf{p}_2 , \mathbf{p}_3 \rangle
\vert {\partial (\mathbf{p}_1 , \mathbf{p}_2 , \mathbf{p}_3)
\over \partial ( \mathbf{P} , \mathbf{q}_{k}, \mathbf{k}_{ij})
 }\vert^{1/2}.  
\label{eq:b.40a}
\end{equation}
The above definitions
imply the desired transformation property
\begin{equation}
{\cal U}_0 (\Lambda ,a) \vert \mathbf{P} , \mathbf{q}_{k}, \mathbf{k}_{ij}
\rangle =
\vert \mathbf{P}', R_w (\Lambda ,P_{\sf W}) \mathbf{q}_k,
R_w (\Lambda ,P_{\sf W}) \mathbf{k}_{ij} \rangle
\sqrt{E_{P_{\sf W}}' \over E_{P_{\sf W}} }  e^{i P_{\sf W}' \cdot a}
\label{eq:b.41}
\end{equation}
where $P'=\Lambda P$. 

The operators $\mathbf{q}_i$ and $\mathbf{k}_i$ are functions of the 
single particle momenta, and are thus defined on states with 
any total momenta, not just on three-body rest states.  This is 
similar to the mass operator, which is also defined on states of 
any total momentum. 

Next we discuss the structure of the mass operators that will be used in the 
two and three-body problems.
The mass operators $m_{0,ij}$ and $M_0$ for the non-interacting two- and
three-body systems can be expressed in terms of the operators
$\tilde{\mathbf{k}}_{ij}$, $\mathbf{k}_{ij}$,  and $\mathbf{q}_{k}$ as
\begin{equation}
m_{0,ij} = \sqrt{m_i^2 + \tilde{\mathbf{k}}_{ij}^2 } +
\sqrt{m_j^2 + \tilde{\mathbf{k}}_{ij}^2 }
\label{eq:b.42}
\end{equation}
and
\begin{equation}
M_0 = \sqrt{m_{0ij}^2 + \mathbf{q}_{k}^2 } +
\sqrt{m_k^2 + \mathbf{q}_{k}^2 }
\label{eq:b.43}
\end{equation}
where $m_{0,ij}$ in (\ref{eq:b.43}) replaces
$\tilde{\mathbf{k}}_{ij}$  by ${\mathbf{k}}_{ij}$.

When two-body interactions are added to $m_{0,ij}$ the interacting
two-body mass operator becomes:
\begin{equation}
m_{ij} = m_{0,ij} + \tilde{v}_{ij}
\label{eq:b.44}
\end{equation}
where in this basis (\ref{eq:b.24}) becomes
\begin{equation}
\langle \mathbf{p}_{ij}, \tilde{\mathbf{k}}_{ij} \vert
\tilde{v}_{ij}
\vert \mathbf{p}_{ij}\,' , \tilde{\mathbf{k}}_{ij}\,' \rangle
=
\delta (\mathbf{p}_{ij} - \mathbf{p}_{ij}\,')
\langle \tilde{\mathbf{k}}_{ij} \Vert v_{ij} \Vert
\tilde{\mathbf{k}}_{ij}\,' \rangle .
\label{eq:b.45}
\end{equation}

Cluster properties determine how the two-body interactions 
enter the three-body mass operator.  In order to obtain a three-body 
scattering operator that clusters properly, it is enough
to replace
$\tilde{\mathbf{k}}_{ij}$ by $ \mathbf{k}_{ij}$ in the 
two-body interaction and include the modified two-body interaction 
in the three-body mass operator as follows:  
\begin{equation}
M_{ij} = \sqrt{(m_{0,ij}+v_{ij})^2 + \mathbf{q}_{k}^2 } +
\sqrt{m_k^2 + \mathbf{q}_{k}^2 }
\label{eq:b.46}
\end{equation}
where 
\begin{equation}
\langle \mathbf{P}, \mathbf{q}_k, \mathbf{k}_{ij} \vert
v_{ij}
\vert \mathbf{P}\,', \mathbf{q}_k\,', \mathbf{k}_{ij}\,' \rangle
=
\delta (\mathbf{P} - \mathbf{P}\,') \delta
(\mathbf{q}_k- \mathbf{q}_k\,')
\langle \mathbf{k}_{ij} \Vert v_{ij} \Vert \mathbf{k}_{ij}\,' \rangle
\label{eq:b.47}
\end{equation}
and the functional form of the
reduced kernel $\langle
\mathbf{k}_{ij} \Vert v_{ij} \Vert \mathbf{k}_{ij}\,' \rangle$
is identical in (\ref{eq:b.45}) and (\ref{eq:b.47}), and it
must be a rotationally invariant function of its arguments
$\tilde{\mathbf{k}}_{ij}$ and $\tilde{\mathbf{k}}_{ij}\,'$
(resp. ${\mathbf{k}}_{ij}$ and ${\mathbf{k}}_{ij}\,'$) .

The interacting three-body mass operator is then defined by
\begin{equation}
M:= M_0 + V_{12} + V_{23} + V_{31}
\label{eq:b.48}
\end{equation}
where the two-body interactions embedded in the three-particle Hilbert space
\cite{fch} are
\begin{equation}
V_{ij} := M_{ij} -M_0
= \sqrt{(m_{0,ij}+v_{ij})^2 + \mathbf{q}_{k}^2 }
- \sqrt{m_{0,ij}^2 + \mathbf{q}_{k}^2 }.
\label{eq:b.49}
\end{equation}
The sum of the interactions is consistent with the constraint
(\ref{eq:b.24}) since each of the two-body interactions $V_{ij}$ in
(\ref{eq:b.48}) is consistent with (\ref{eq:b.24}). The dynamical
problem is to find eigenstates of $M$ in the basis (\ref{eq:b.40a}).
The technical challenge for the numerical solutions of the three
body-problem is the computation of the embedded two-body interactions,
$V_{ij}$, which requires computing functions of the non-commuting 
operators $m_{0,ij}$ and $v_{ij}$.

While this dynamical model leads to an $S$-matrix that satisfies
cluster properties, the constructed unitary 
representation of the Poincar\'e group 
only clusters properly when $\mathbf{P}=\mathbf{0}$.  Since the
$S$-matrix is Poincar\'e invariant, this is sufficient for computing all 
bound state and three-body scattering observables, however additional 
corrections are required if the three-body eigenstates are
used to compute electromagnetic observables.

\section{Relativistic Scattering Theory} 

Our interest in this paper is the computation of scattering cross
sections for two-body elastic scattering and breakup reactions in
Poincar\'e invariant quantum mechanics.  The formulation of the
scattering theory using dynamical mass operators for the Poincar\'e
group is outlined below.  For a more complete discussion see
\cite{bkwp}.

The multichannel scattering matrix is calculated using the standard
formula
\begin{equation}
S_{\alpha \beta} = \langle \Psi_\alpha^+ (0) \vert \Psi_\beta^- (0)
\rangle
\label{eq:b.50}
\end{equation}
where $\beta$ is the initial deuteron-nucleon channel and 
$\alpha$ is either a deuteron-nucleon or three nucleon channel.

The scattering states that appear in Eq.~(\ref{eq:b.50}) 
are defined to agree with states of non-interacting particles
long before or long after the collision
\begin{equation}
\lim_{t \to \pm \infty} \Vert e^{-iHt} \vert \Psi^{\pm}_\alpha (0) \rangle
- e^{-i H_\alpha t} \vert \Phi^{\pm}_\alpha (0) \rangle \Vert =0 .
\label{eq:b.51}
\end{equation}
In this paper the $\pm$ on the scattering states and wave operators indicates
the direction of the time limit ($-=$past/$+=$future), which is opposite 
to the sign of $i\epsilon$.

In the breakup channel $\vert \Phi^{+}_\alpha (0)\rangle $ is a normalizable 
Hilbert space vector.
In the nucleon-deuteron channels 
$\vert \Phi^{\pm}_{\alpha} (0) \rangle $ has the form 
\begin{equation}
\langle \mathbf{P}, \mathbf{q}_i , \mathbf{k}_{jk} \vert 
\Phi^{\pm}_{\alpha} (0) \rangle = 
\phi_D(\mathbf{k}_{jk})  f( \mathbf{q}_i ,\mathbf{P})
\end{equation} 
where $\phi_D(\mathbf{k}_{jk})$ is the deuteron wave function and 
$f( \mathbf{q}_i ,\mathbf{P})$ is a unit normalized wavepacket 
describing the state of a free deuteron and nucleon at time zero.

The asymptotic and interacting scattering states are related by the
multichannel wave operators
\begin{equation}
\vert \Psi_\alpha^{\pm} (0) \rangle =
\Omega_{\alpha\pm }   (H, H_\alpha)
\vert \Phi_\alpha^{\pm} (0) \rangle
\label{eq:b.53}
\end{equation}
where the multichannel wave operators are defined by the strong limits
\begin{equation}
\Omega_{\alpha \pm} = \lim_{t \to \pm \infty}
e^{i H t } e^{-i H_\alpha t} 
\label{eq:b.54}
\end{equation}
on channel states.
The multichannel scattering operator can then be expressed in terms of the
wave operators as
\begin{equation}
\hat{S}_{\alpha \beta} =
\Omega_{\alpha + } ^{\dagger}  (H, H_\alpha) \Omega_{\beta -}
(H, H_\beta) .
\label{eq:b.55}
\end{equation}

\noindent
In the three-body breakup channel, $\alpha=0$,
\begin{equation}
H_\alpha = H_0 = \sqrt{M_0^2 + \mathbf{P}^2}  .
\label{eq:b.56}
\end{equation}
In channels, $\alpha= (ij)$, with an incoming or outgoing deuteron,
\begin{equation}
H_\alpha = H_{ij} = H_0 + V_{H \alpha},
\label{eq:b.57}
\end{equation}
where 
\begin{equation}
V_{H \alpha}= \sqrt{M_{ij}^2 + \mathbf{P}^2} - \sqrt{M_0^2 + \mathbf{P}^2}
\label{eq:b.58}
\end{equation}
is the interaction between the nucleons in the deuteron in the
three-body Hamiltonian.  We use the notation $M_\alpha$ to denote
$M_0$ for the breakup channel or $M_{ij}$ for the nucleon-deuteron
channel.  In the non-relativistic case the Hamiltonian, which 
generates time evolution in the asymptotic conditions, is
normally replaced by the Casimir Hamiltonian, $h$.  This can be done
because $\mathbf{P}^2 / 2 M_g$ appears linearly in both the
interacting and non-interacting Hamiltonians and commutes with the
interactions.  In the relativistic case the interaction in the
Hamiltonian is different from the interaction in the mass operator,
and the kinetic energy enters the mass non-linearly.
In the relativistic case the wave operators can still be expressed
directly in terms of the mass operators.  The justification is the
Kato-Birman invariance principle \cite{cg,baum} which implies that $H$
and $H_\alpha$ can be replaced by a large class of admissible
functions of $H$ and $H_\alpha$ in the wave operators; specifically
\begin{equation}
M= \sqrt{H^2 - \mathbf{P}^2}
\label{eq:b.59}
\end{equation}
is in the class of admissible functions.  This gives
\begin{equation}
\Omega_{\alpha \pm} = \lim_{t \to \pm \infty}
e^{i M t } e^{-i M_\alpha t} =
\lim_{t \to \pm \infty}
e^{i H t } e^{-i H_\alpha t}
\label{eq:b.60}
\end{equation}
which leads to an expression for the
multichannel scattering operator \cite{fcwp} expressed directly in terms
of the mass operators:
\begin{equation}
S_{\alpha \beta} = \lim_{\tau,\tau' \to \infty}
e^{i M_\alpha \tau} 
e^{-iM (\tau+\tau')}
e^{i M_\beta \tau'} .
\label{eq:b.61}
\end{equation}
If these limits are computed in channel mass eigenstates 
$\vert \alpha \rangle$ and
$\vert \beta \rangle$ of
$M_{\alpha}$ and $M_{\beta}$ the result is
\begin{equation}
\langle \alpha  \vert S \vert \beta \rangle =
\langle \alpha \vert \beta \rangle 
- 2 \pi i \delta ({\sf W}_{\alpha}- {\sf W}_{\beta})
\langle \alpha \vert T^{\alpha \beta}({\sf W}_\alpha +i0^+) \vert
\beta \rangle  ,
\label{eq:b.62}
\end{equation}
where
\begin{equation}
T^{\alpha \beta} (z) = V^{\beta } +
V^{\alpha } (z-M)^{-1} V^{\beta} ,
\label{eq:b.63}
\end{equation}
and
\begin{equation}
V^{\alpha} = M - M_{\alpha} = M-M_{ij} 
\label{eq:b.64}
\end{equation}
for two cluster $n-d$ channels and 
\begin{equation}
V^{\alpha} = M - M_{0}  
\label{eq:b.65}
\end{equation}
for the breakup channel.
Here ${\sf W}_\alpha$ and ${\sf W}_\beta$ are the eigenvalues of
$M_{\alpha}$ and $M_{\beta}$ in the channel eigenstates $\vert \alpha
\rangle$ and $\vert \beta \rangle$.  The first term in
Eq.~(\ref{eq:b.62}) is identically zero if the states $\vert \alpha
\rangle$ and $\vert \beta \rangle$ correspond to different scattering
channels.

Compared to the standard expression that is based on using the
Hamiltonian, in Eq.~(\ref{eq:b.63}) the interactions are
expressed as differences of mass operators rather than Hamiltonians,
the resolvent of the Hamiltonian is replaced by the resolvent of the
mass operator and the energy conserving delta function is replaced
with an invariant mass conserving delta function.

The translational invariance of the interaction
(\ref{eq:b.64}) requires that
\begin{equation}
\langle \mathbf{P} , \cdots \vert
T^{\alpha \beta} (z) \vert \cdots ,  {\bf P}\,' \rangle =
\delta (\mathbf{P}- \mathbf{P}\,')
\langle \cdots \Vert T^{\alpha \beta} (z) \Vert \cdots \rangle .
\label{eq:b.66}
\end{equation}
Given the momentum conserving delta function, the product of the 
momentum and mass conserving delta functions can be replaced by a 
four-momentum conserving delta function and a Jacobian:
\begin{equation}
\delta ({\sf W}_\alpha - {\sf W}_{\beta} )\delta (\mathbf{P}_{\alpha} - 
\mathbf{P}_{\beta} ) 
=
\delta^4 ({\sf P}_\alpha - {\sf P}_{\beta}) \left| {dE \over dM}\right| =
\delta^4 ({\sf P}_\alpha - {\sf P}_\beta)
\left| {{\sf W}_\alpha  \over {\sf E}_\alpha}\right|
\label{eq:b.66b}
\end{equation}
where ${\sf E}_\alpha=\sqrt{{\sf W}_{\alpha}^2 + \mathbf{P}^2}$.

The representation (\ref{eq:b.62}) of the scattering matrix can be 
used to calculate
the cross section.  The relation between the scattering matrix and the
cross section is standard and can be derived using standard
methods, such as the ones used by Brenig and Haag in \cite{bh}.  
The only modification is
that in the usual expression relating the cross section to the
transition operator, the transition operator is the coefficient of
$-(2 \pi)i \delta (E_\alpha -E_\beta)$.  Thus to compute the cross section it
is enough to use the standard relation between $T$ and $S$ with the
channel transition operator being replaced by
\begin{equation}
\left| {{\sf W}_\alpha  \over {\sf E}_\alpha}\right| \times 
\langle \cdots \Vert
T^{\alpha \beta} (z) \Vert \cdots  \rangle
\label{eq:b.67}
\end{equation}

The resulting expression for the differential
cross section for elastic scattering is given by :
\begin{equation}
{d\sigma} =
{(2\pi)^{4}\over v_{nd}'}
\left|  \langle \mathbf{p}_d, \mathbf{p}_n
\Vert T^{\alpha \beta} \Vert
\mathbf{p}_d',  \mathbf{p}_n'  \rangle \right| ^2
{{\sf W}_\alpha^{ 2}  \over {\sf E}^{2}_\alpha}
\;
\delta^4 \left( p_d + p_n  - {p}_d' - {p}_n' \right) \;
d\mathbf{p}_n d\mathbf{p}_d  .
\label{eq:b.70}
\end{equation}
and for breakup reactions the formula is replaced by 
\begin{equation}
{d\sigma} =
{(2\pi)^{4}\over v_{nd}' 3!}
\left|  \langle \mathbf{p}_1, \mathbf{p}_2, \mathbf{p}_3 
\Vert T^{\alpha \beta} \Vert
\mathbf{p}_d',  \mathbf{p}_n'  \rangle \right| ^2
{{\sf W}_\alpha^{ 2}  \over {\sf E}^{ 2}_\alpha}
\;
\delta^4 \left( p_1 + p_2 + p_3 - {p}_d' - {p}_n' \right) \;
d\mathbf{p}_1 d\mathbf{p}_2  d\mathbf{p}_3  .
\label{eq:b.71}
\end{equation}
These relations are normally given in terms of single particle 
momenta while the transition matrix elements are evaluated 
in terms of the Poincar\'e-Jacobi momenta.  The transformation 
relating  these representations involves some Jacobians.  These are discussed 
in the sections on calculations.

Except for the factor ${\sf W}_\alpha^2 / {\sf E}^2_\alpha$,
Eq.~(\ref{eq:b.70}) is identical to the corresponding non-relativistic
expression.  The additional factor of ${\sf W}_\alpha^2 / {\sf
E}^2_\alpha$ arises because we have chosen to calculate the transition
operator using the mass operator instead of the Hamiltonian.  The
differences in these formulas with standard formulas are (1) that the
transition operator is constructed from the difference of the mass
operators with and without interactions and (2) the appearance of the
additional factor of ${\sf W}_\alpha^2 / {\sf E}^2_\alpha$ which
corrects for the modified transition operator.  This factor becomes 1
when $\mathbf{P}=0$.

The differential cross section $d \sigma$ is invariant.  Equations  
(\ref{eq:b.70}) and (\ref{eq:b.71}) can be expressed in a manifestly
invariant form.  The relation to the standard expression of the invariant
cross section using conventions of the particle data book \cite{pdg}
is derived in the appendix where we also outline the proof of
Eq.~(\ref{eq:b.62}).


\section{Integral Equations}

The dynamical problem is to compute the three-body transition 
operators $T^{\alpha \beta}(z)$ that appear in
Eqs.~(\ref{eq:b.70}) and (\ref{eq:b.71}) and use these to 
calculate the cross sections.
It is useful to replace the transition operators (\ref{eq:b.63}) by 
the on-shell equivalent AGS \cite{alt} transition operators:
\begin{equation}
U^{\alpha \beta}(z) := \bar{\delta}^{\alpha \beta}(z-M_\alpha)
+ T^{\alpha \beta}(z).
\label{eq:c.1}
\end{equation}
When $z$ is put on the energy shell and evaluated on the channel 
eigenstate for the channel $\alpha$ the first term of (\ref{eq:c.1})
vanishes.  The AGS operators 
are solutions of the integral equation
\begin{equation}
U^{\alpha \beta}(z) := \bar{\delta}^{\alpha \beta}(z-M_0) +
\sum_\gamma \bar{\delta}^{\alpha \gamma} V_\gamma (z-M_\gamma)^{-1}
U^{\gamma \beta}(z),
\label{eq:c.2}
\end{equation}
where $V_\gamma:=M_\gamma - M_0 = V_{ij}$ are the embedded two-body
interactions given in (\ref{eq:b.49}), and the sum is over the 
three two cluster configurations, $1=(1;23),2=(2;31),3=(3;12)$,
with  each cluster labeled by the index $\gamma$.

When the particles are identical this coupled
system can be replaced by an equation for a single amplitude,
\begin{equation}
U (z) = P (z-M_0) + P V_{1}(z-M_1)^{-1} U(z),
\label{eq:c.6}
\end{equation}
where we chose without loss of generality to single out 
the configuration (1;23). In this case the
permutation operator $P$ is given by  $P=P_{12}P_{23}+P_{13}P_{23}$.
This solution can be used to generate the breakup amplitude
\begin{equation}
U_0= (I+P) V_1 (z-M_1)^{-1} U(z).
\label{eq:c.7}
\end{equation}
The AGS operators $U(z)$
and $U_0(z)$ can be expressed in terms of the solution $T(z)$ of the
symmetrized
Faddeev equations
\begin{eqnarray}
U(z) &=& P (z-M_0) + PT(z) \nonumber \\
U_0(z) &=& (1+P) T(z) ,
\label{eq:2.3}
\end{eqnarray}
where $T(z)$ is the solution to
\begin{equation}
T(z)  = T_1(z) P +T_1(z) P (z-M_0)^{-1}  T(z) .
\label{eq:2.4}
\end{equation}
where the operator $T_1(z)$ is the two-body transition operator
embedded in the three-particle Hilbert space 
defined as the solution to
\begin{equation}
T_1(z)=V_1+V_1 (z-M_0)^{-1}T_1(z) ,
\label{eq:2.5}
\end{equation}
where $V_1=V_{23} =  M_{23}-M_0$.

The first order calculation that we perform in this paper
is defined by keeping the first term of the multiple scattering series
generated by (\ref{eq:2.4}):
\begin{equation}
T(z)  \to T_1(z) P
\end{equation}
\begin{equation}
U(z) \to P (z-M_0) + PT_1 (z)P \qquad
U_0(z) \to (1+P) T_1(z)P ,
\end{equation}

Because the embedded interactions $V_\gamma$ (\ref{eq:b.49}) in the 
AGS equations are operator valued functions of the non-commuting operators 
$m_{0,ij}$ and $v_{ij}$, their computation,  given $v_{ij}$ as input, 
presents additional computational 
challenges 

To compute the kernel 
note that it can 
be expressed as 
\begin{equation}
V_\gamma (z-M_\gamma)^{-1} = T_\gamma (z)  (z-M_0)^{-1} 
\label{eq:c.3}
\end{equation}
where
\begin{equation}
T_\gamma (z) = V_\gamma + V_{\gamma}  (z-M_\gamma)^{-1} V_\gamma 
\label{eq:c.4}
\end{equation}
In this paper we compute this kernel using a method that exploits the 
relation between the two-body transition operator and the operator 
(\ref{eq:c.4}).
Because $M_{ij}$ and $m_{ij}$ have the same eigenvectors 
it follows that \cite{keipo} 
\begin{equation}
\langle \mathbf{P}', \mathbf{q}' , \mathbf{k}' \vert 
T_{\gamma}(z')  (z'-M_0)^{-1} \vert \mathbf{P}, \mathbf{q} , \mathbf{k} \rangle =
\delta (\mathbf{P}\,'-\mathbf{P})\delta (\mathbf{q}\,'-\mathbf{q})
\left ( {m_{0,ij} (\mathbf{k}') + 
m_{0,ij} ( \mathbf{k})
\over
M_0 (\mathbf{q} , \mathbf{k}') + 
M_0 (\mathbf{q} , \mathbf{k})}
\right ) 
{\langle \mathbf{k}^{\prime} \vert
t_{ij} ( \tilde{z}') \vert   \mathbf{k}^{} 
\rangle \over  
z'-M_0(\mathbf{q} , \mathbf{k})} 
\label{eq:c.8}
\end{equation}
where
\begin{equation}
z'=M_0 (\mathbf{q} , \mathbf{k}\,' ) + i0^+ \qquad 
\tilde{z}'= m_{0ij} (\mathbf{k}\,' ) + i0^+ .
\label{eq:c.9}
\end{equation}
In the AGS equation, this kernel is needed for all values of 
$z$, while equation (\ref{eq:c.8}) only holds for $z=z'$.
The kernel for an arbitrary $z$ can be computed 
by using the first resolvent
equation which leads to integral equation  
\begin{equation}
T_\gamma (z) = 
T_\gamma (z')+
T_\gamma (z)G_0(z) (z'-z)G_0(z') T_\gamma (z')
\label{eq:c.10}
\end{equation}
where $G(z)= (z-M_0)^{-1}$
which can be used to calculate $T_\gamma(z)$ from 
$T_\gamma (z')$ for all $z\not= z'$.

\section{Relativistic Formulation of Three-Body Scattering}

In the scattering of three particles interacting with spin independent 
interactions, there are two global observables,
the total cross section for elastic scattering, $\sigma_{el}$, and the
total cross section for breakup, $\sigma_{br}$.  These can be computed
using (\ref{eq:b.70}) and (\ref{eq:b.71}).  In this section we 
discuss the kinematic 
relations needed to compute these quantities in more detail. 

If we replace the transition operators by 
the corresponding symmetrized AGS operators,  use the identities
\begin{equation} 
\Pi_d= \int  \vert {\mathbf P} ,{\mathbf q},\varphi_d \rangle  
d{\mathbf P}  d{\mathbf q} 
\langle {\mathbf P} ,{\mathbf q} ,\varphi_d \vert  =
\int 
\vert {\mathbf p}_d ,{\mathbf p}_n ,\varphi_d \rangle  
d{\mathbf p}_d d{\mathbf p}_n  
\langle {\mathbf p}_d ,{\mathbf p}_n ,\varphi_d \vert  
\end{equation}
and
\begin{equation}
I= \int 
\vert {\mathbf P} ,{\mathbf k} ,{\mathbf q} \rangle  
d{\mathbf P} d{\mathbf k} d{\mathbf q} 
\langle {\mathbf P} ,{\mathbf k} ,{\mathbf q} \vert  =
\int 
\vert {\mathbf p}_1 ,{\mathbf p}_2 ,{\mathbf p}_3 \rangle  
d{\mathbf p}_1 d{\mathbf p}_2 d{\mathbf p}_3 
\langle {\mathbf p}_1 ,{\mathbf p}_2 ,{\mathbf p}_3 \vert ,  
\end{equation}
and evaluate the initial state and $v_{bt}$ in the center of 
momentum frame, (\ref{eq:b.70}) becomes
\begin{equation}
\sigma_{el} = \frac{(2\pi)^4}{v_{bt}} \int d\mathbf{q}\,
\delta({\sf W}_f-{\sf W}_i)|\langle
\varphi_d , {\bf {q}} \Vert U \Vert\varphi_d, {\bf q_0}\rangle|^2
\label{eq:2.1}
\end{equation}
for elastic scattering and
\begin{equation}
\sigma_{br} = \frac{1}{3}\frac{(2\pi)^4}{v_{bt}} \int d\mathbf{q} \;
d \mathbf{k}
\delta({\sf W}_f-{\sf W}_i) |\langle{\bf k},{\bf q}
\Vert U_0\Vert \varphi_d , {\bf
q_0}\rangle|^2.
\label{eq:2.2}
\end{equation}
for breakup.

Here ${\sf W}_i ({\sf W}_f)$ are the invariant masses eigenvalues of
the initial (final) state, ${\bf q_0}$ is the Poincar\'e-Jacobi momentum
between the projectile and the target, and {\bf k} and {\bf q} the
Poincar\'e-Jacobi momenta for a given pair and spectator defined in
the previous section.   The permutation operator in
(\ref{eq:c.6}) only includes three of the six permutations of the
three particles; the other three independent permutations are related
by an additional transposition that interchanges the constituents of
the deuteron, which is already symmetrized.  This accounts for
replacement of the statistical factor $1/3!$ in (\ref{eq:b.70}) 
by the factor of $1/3$ in
(\ref{eq:2.2}).

Using relativistic kinematical relations the integral over
$\vert \mathbf{q} \vert$ in Eq~(\ref{eq:2.1}) can be done using 
the invariant mass conserving delta functions with the result
\begin{equation}
\sigma_{el}= (2\pi)^4 \int d\Omega \frac{E_n^2 (q_0) E_d^2(q_0)}
{{\sf W}^2}
|\langle \varphi_d, {\bf {\hat q}} q_0|U|\varphi_d, {\bf q_0}\rangle|^2.
\label{eq:2.6}
\end{equation}
The quantities ${\sf W}$ and $\mathbf{q}_0$ are determined 
by the 
laboratory kinetic energy $E_{lab}$ of the incident nucleon.
First note 
\begin{equation}
{\sf W}^2 = ( m+m_d )^2 + 2m_d E_{lab}.
\label{eq:2.7}
\end{equation}
The nucleon rest mass is given by $m$, the rest mass of the deuteron
is $m_d = 2m - \varepsilon_d$, where $\varepsilon_d$ is the deuteron
binding energy.  The Poincar\'e Jacobi momentum between projectile and
target, ${\bf q_0}$, is related to $E_{lab}$ by
\begin{equation} {\bf q_0^2} = \frac{m_d^2 E_{lab}}{{\sf W}^2} \left(
{E_{lab} + 2m} \right) \label{eq:2.8} \end{equation} In the
nonrelativistic case the phase space factor under the integral of
Eq.~(\ref{eq:2.6}) reduces to $(2m/3)^2$.

\begin{figure}
\begin{center}
\includegraphics[width=10cm]{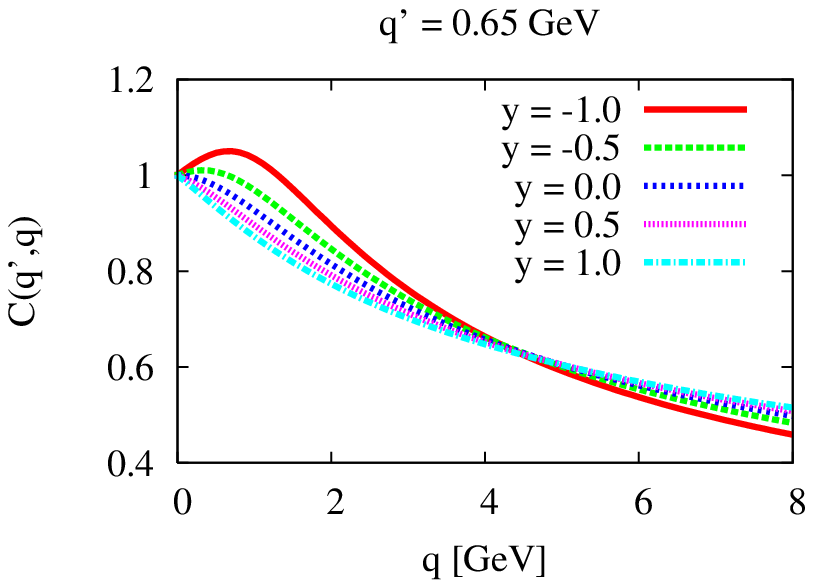}
\end{center}
\caption{(Color online) The momentum and angle dependence of the function $C({\mathbf
q}',{\mathbf q})$
from Eq.~(\ref{eq:2.13}) at fixed momentum $q'$~=~0.65~GeV.
\label{fig1}}
\end{figure}

\begin{figure}
\begin{center}
\includegraphics[width=10cm]{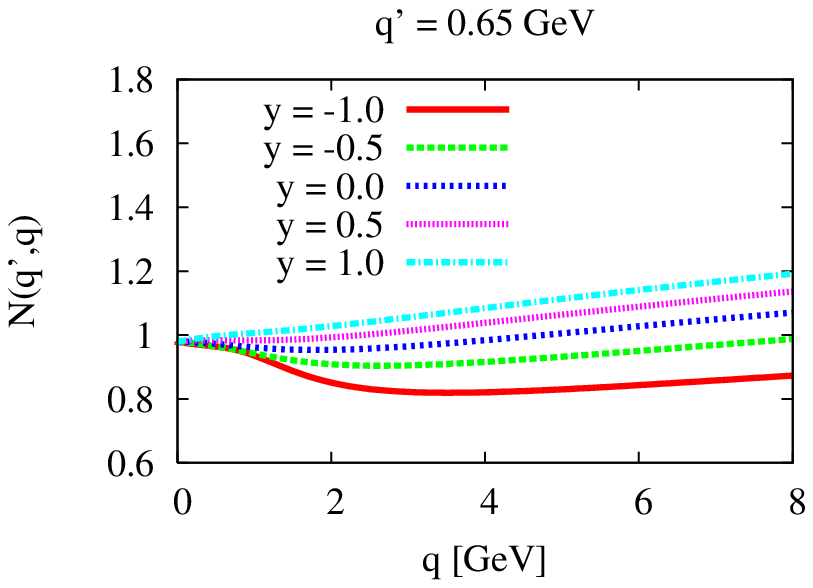}
\end{center}
\caption{(Color online) The momentum and angle dependence of the function $N({\mathbf
q}',{\mathbf q})$
from Eq.~(\ref{eq:2.12}) at fixed momentum $q'$~=~0.65~GeV.
\label{fig2}}
\end{figure}

It is also necessary to compute the transition matrix elements that 
appear in (\ref{eq:2.1}) and (\ref{eq:2.2}).
The momenta of the three particles can be labeled either by 
single-particle momenta
${\bf p_1}$, ${\bf p_2}$, and ${\bf p_3}$, or the
total momentum $\mathbf{P}$ and the relativistic Poincar\'e-Jacobi
momenta  of Eqs.~(\ref{eq:b.37}) and (\ref{eq:b.38}) with the
$p_i$ replaced by $q_i$.   The explicit relations
between the three-body Poincar\'e Jacobi momenta are
\begin{eqnarray}
{\bf q} & \equiv & {\bf q_i} = - ({\bf q_j} + {\bf q_k} ) \nonumber \\
{\bf k} & \equiv & {\bf k_i} =
{\bf k_{jk}} = \frac{1}{2}\left({\bf
q_j}-{\bf q_k} \right) - \frac{1}{2}\left( {\bf q_j}+{\bf q_k} \right)
\left(\frac{E_j-E_k}{E_j+E_k + \sqrt{(E_j+E_k)^2 -({\bf q_j}+{\bf
q_k})^2}}\right),
\label{eq:2.9}
\end{eqnarray}
where $E_i\equiv E({\bf q_i})=\sqrt{m^2 +{\bf q_i}^2}$. 
For nonrelativistic kinematics the second term in {\bf k}, being
proportional to the total momentum of the pair $(j,k)$, vanishes.  In
addition, the transformation from the single particle
momenta ${\bf p_i}$ to the
Poincar{\'e}-Jacobi momenta has a Jacobian given by
\begin{equation}
|{\bf p_1}, {\bf p_2}, {\bf p_3}\rangle  = 
\left|\frac{\partial({\bf P},{\bf k},{\bf q})}
{\partial({\bf p_1},{\bf p_2},{\bf p_3})}\right|^{1/2}
|{\bf P}, {\bf k}, {\bf q}\rangle 
\end{equation}
where for ${\bf P}=\mathbf{0}$ the Jacobian becomes
\begin{equation}
\left|\frac{\partial({\bf P},{\bf k},{\bf q})} {\partial({\bf
p_1},{\bf p_2},{\bf p_3})}\right|^{1/2}_{\vert_{{\bf P}=\mathbf{0}}} = \left( \frac{\sqrt{(E({\bf
q_2})+E({\bf q_3}))^2 -{\bf q}^2} \;(E({\bf q_2})+E({\bf q_3}))}
{4E({\bf q_2})E({\bf q_3})} \right) ^{1/2} 
\equiv  {\hat n}\left({\bf q};{\bf
q_2}{\bf q_3}\right) .
\label{eq:2.10}
\end{equation}
In the above expression we chose without loss of generality particle 1
as spectator.  The difference between the relativistic and
non-relativistic Jacobi momenta in Eqs.~(\ref{eq:b.37}) and
(\ref{eq:b.38}) are relevant for the calculation of the permutation operator
$P$ in Eqs.~(\ref{eq:2.3}) and (\ref{eq:2.4}). The matrix elements of
the permutation operator are then explicitly calculated as
\begin{eqnarray}
\langle {\bf k'},{\bf q'}|P|{\bf k},{\bf q}\rangle &=& N({\mathbf
q}',{\mathbf q}) \Bigg[ \delta ({\mathbf k}'-{\mathbf q} -
\frac{1}{2}{\mathbf q}' \: C({\mathbf q},{\mathbf q}') )
\delta({\mathbf k}+{\mathbf q}' +\frac{1}{2}{\bf q} \: C(({\bf
q'},{\bf q})) \nonumber \\ &+& \delta({\mathbf k}'+{\mathbf q}
+\frac{1}{2}{\bf q}' \: C({\bf q},{\bf q}')) \delta({\mathbf
p}-{\mathbf q}' -\frac{1}{2}{\bf q} \: C(({\bf q'},{\bf q}))\Bigg],
\label{eq:2.11}
\end{eqnarray}
where the function $N({\mathbf q}',{\mathbf q})$ contains the
product of two Jacobians and reads
\begin{eqnarray}
N(\mathbf{q},\mathbf{q}') &=& \frac{\sqrt{E(\mathbf{q}) +
E(\mathbf{q} + \mathbf{q}')} \; \sqrt{E(\mathbf{q}') + E(\mathbf{q} +
\mathbf{q}')}} {4E(\mathbf{q} + \mathbf{q}') } \nonumber\\ 
& & \times
\frac{\sqrt[4]{(E(\mathbf{q}) + E(\mathbf{q} + \mathbf{q}'))^2 -
\mathbf{q}'^2} \;\; \sqrt[4]{(E(\mathbf{q}') + E(\mathbf{q} + \mathbf{q}'))^2
- \mathbf{q}^2}} {\sqrt{E(\mathbf{q})E(\mathbf{q}')}} .
\label{eq:2.12}
\end{eqnarray}
and the function $C({\mathbf q},{\mathbf q}')$ is calculated as
\begin{equation}
C({\mathbf q}',{\mathbf q})
= 1 + \frac{E({\mathbf q}')-E({\mathbf q}'+{\mathbf q})}
{E({\mathbf q}')+E({\mathbf q}'+{\mathbf q})
+ \sqrt{(E({\mathbf q}')+E({\mathbf q}'+{\mathbf q}))^2 -{\mathbf q}^2} } .
\label{eq:2.13}
\end{equation}
These permutation operators, which change the order of coupling, are
essentially Racah coefficients for the Poincar\'e group.  In the
nonrelativistic case the functions $N({\mathbf q}',{\mathbf q})$ and
$C({\mathbf q}',{\mathbf q})$ both reduce to the constant 1 and have
the relatively compact form of the matrix elements of $P$ given in
e.g. \cite{bound3d,scatter3d}.  Since both functions depend on
magnitudes as well as angles between the momentum vectors, the 3D
formulation is very appropriate for our relativistic calculations. In
order to illustrate the momentum and angle dependence we display in
Fig.~\ref{fig1} the function $C({\mathbf q}',{\mathbf q})$ for a given
value of $|{\bf q'}|$~=~0.65~GeV as function of $|{\bf q}|$ and
several values of the angle $y={\bf \hat q'}\cdot {\bf \hat q}$. In
general the values of $C$ drop below 1 as $q$ increases. The angle
dependence is strongest for small $q$, where for ${\bf \hat q'}\cdot
{\bf \hat q} = -1$ the function is larger than 1.  For the same fixed
value of $|{\bf q'}|$ we display the function $N({\mathbf q}',{\mathbf
q})$ in Fig.~\ref{fig2}. Here we see a slowly varying dependence on
the momentum $|{\bf q}|$ and a strong angle dependence. For small
angles ($y=1$) the function $N({\bf q}',{\bf q}) $ is larger than 1,
whereas for large angles ($y=-1$) it is reduced from 1 by as much as
20\%.

In matrix form the Faddeev equation, Eq.~(\ref{eq:2.4}),
has the form
\begin{equation}
\langle \mathbf{k}, \mathbf{q} \Vert T \Vert \varphi_d, \mathbf{q}_0 \rangle =
\langle \mathbf{k}, \mathbf{q} \Vert T_1 P 
\Vert \varphi_d, \mathbf{q}_0 \rangle +
\langle \mathbf{k}, \mathbf{q} \Vert T_1 P(z-M_0)^{-1} 
T \Vert \varphi_d, \mathbf{q}_0
\rangle .
\label{eq:2.14}
\end{equation}
Since at this stage we are only carrying out first order calculations,
we only need to consider the first term. Explicitly this is given as
\begin{eqnarray}
\langle \mathbf{k}, \mathbf{q} \Vert T({\sf W}) \Vert \varphi_d, \mathbf{q}_0 \rangle
&=& \langle \mathbf{k}, \mathbf{q} \Vert T_1(\mathbf{q},\varepsilon ) P \Vert
\varphi_d ,\mathbf{q}_0 \rangle \nonumber \\ &=& \int {
d \mathbf{k}'d\mathbf{q}'d\mathbf{k}''d\mathbf{q}''
\langle \mathbf{k}, \mathbf{q} \Vert
T_1(\mathbf{q},\varepsilon) \Vert \mathbf{k}', \mathbf{q}' \rangle \langle
\mathbf{k}', \mathbf{q}' \Vert P \Vert \mathbf{k}'', \mathbf{q}'' \rangle
\langle \mathbf{k}'' ,\mathbf{q}'' | \varphi_d, \mathbf{q}_0 \rangle }
\nonumber \\ &=& \int d{\mathbf k}'d\mathbf{q}'d\mathbf{k}''d\mathbf{q}''
\; T_1(\mathbf{k},\mathbf{k}',\mathbf{q}; \varepsilon
)\delta(\mathbf{q}'-\mathbf{q}) \varphi_d (\mathbf{k}'')
\delta(\mathbf{q}''-\mathbf{q}_0) \nonumber \\ & & \times
N(\mathbf{q}',\mathbf{q}'') [ \delta(\mathbf{k}'-\mathbf{k}
(\mathbf{q}'', -\mathbf{q}'-\mathbf{q}'') )
\delta(\mathbf{k}''+\mathbf{k} (\mathbf{q}',
-\mathbf{q}'-\mathbf{q}'') ) \nonumber \\ & & +
\delta(\mathbf{k}'+\mathbf{k}(\mathbf{q}'',
-\mathbf{q}'-\mathbf{q}''))
\delta(\mathbf{p}''-\mathbf{p}(\mathbf{q}',
-\mathbf{q}'-\mathbf{q}''))] \nonumber \\ 
&=& N(\mathbf{q},\mathbf{q}_0) \; T_s\left(\mathbf{p},\mathbf{q}_0 +
\frac{1}{2}\mathbf{q} \:
C(\mathbf{q}_0,\mathbf{q}),\mathbf{q};\varepsilon \right) \varphi_d
\left(\mathbf{q} + \frac{1}{2}\mathbf{q}_0 \:
C(\mathbf{q},\mathbf{q}_0) \right) .
\label{eq:2.15}
\end{eqnarray}
Here the invariant parametric energy $\varepsilon$ which enters the two-body
t-matrix is given by $\varepsilon = W -\sqrt{m^2+\mathbf{q}^2}$. Since
we consider bosons, the symmetrized two-body transition matrix $T_s$
is given by
\begin{eqnarray}
T_s(\mathbf{k},\mathbf{k}',\mathbf{q};\varepsilon) &=&
T_1(\mathbf{k},\mathbf{k}',\mathbf{q};\varepsilon) +
T_1(-\mathbf{k},\mathbf{k}',\mathbf{q};\varepsilon) \nonumber \\ &=&
T_1(\mathbf{k},\mathbf{k}',\mathbf{q};\varepsilon) +
T_1(\mathbf{k},-\mathbf{k}',\mathbf{q};\varepsilon) .
\label{eq:2.16}
\end{eqnarray}
This expression is the starting point for our numerical calculations
of the transition amplitude in first order. The first step for an
explicit calculation is the selection of independent variables.  Since
we ignore spin and iso-spin dependencies, the matrix element $\langle
{\bf k},{\bf q}\Vert T\Vert \varphi_d, {\bf q_0}\rangle$ is a scalar
function of the variables $\mathbf{k}$ and $\mathbf{q}$ for a given
projectile momentum $\mathbf{q}_{0}$.  Thus one needs 5 variables to
uniquely specify the geometry of the three vectors $\mathbf{k}$,
$\mathbf{q}$ and $\mathbf{q}_{0}$.  We follow here the procedure from
Ref.~\cite{scatter3d} and choose as variables
\begin{equation}
k=|{\mathbf{k}}|,\ q=|{\mathbf{q}}|,\
x_{k}=\hat{{\mathbf{k}}}\cdot\hat{{\mathbf{q}}}_{0},\
x_{q}=\hat{{\mathbf{q}}}\cdot\hat{{\mathbf{q}}}_{0},\ x^{q_{0}}_{kq}=
\widehat{({\mathbf{q}}_{0}\times{\mathbf{q}})}\cdot
\widehat{({\mathbf{q}}_{0}\times{\mathbf{k}})}.
\label{variables}
\end{equation}
The last variable, $x^{q_{0}}_{kq}$, is the angle between the two
normal vectors of the $\mathbf{k}$-$\mathbf{q}_{0}$-plane and the
$\mathbf{q}$-$\mathbf{q}_{0}$-plane, which are explicitly given by
\begin{eqnarray}
\widehat{({\mathbf{q}}_{0}\times{\mathbf{k}})}&=&\frac{\hat{\mathbf{q}}_{0}
\times\hat{\mathbf{k}}}
{\sqrt{1-(\hat{\mathbf{q}}_{0}\cdot\hat{\mathbf{k}})^{2}}}, \nonumber
\\
\widehat{({\mathbf{q}}_{0}\times{\mathbf{q}})}&=&\frac{\hat{\mathbf{q}}_{0}
\times\hat{\mathbf{q}}}
{\sqrt{1-(\hat{\mathbf{q}}_{0}\cdot\hat{\mathbf{q}})^{2}}}.
\label{normals}
\end{eqnarray}
With these definitions of variables the expression for the transition
amplitude as function of the 5 variables from Eq.~(\ref{variables})
has the same form as its nonrelativistic counterpart, and we can apply
the algorithm developed in Ref.~\cite{scatter3d} for solving it
without partial wave decomposition. The only additional effort is the
evaluation of the functions $C(q_0,q,x_q)$ and 
$N(q_0,q,x_q)$. However, this is conceptually and numerically quite
straightforward, since both functions depend only one angle, $x_q$.


\section{Two-Body Transition Operator and  Relativistic Dynamics}

The kinematic effects related to the use of the relativistic Racah
coefficients have been described in the previous Section. It is left
now, to obtain the transition amplitude of the 2N subsystem embedded
in the three-particle Hilbert space ,
$T_s(\mathbf{k},\mathbf{k}',\mathbf{q};\varepsilon)$, entering
Eq.~(\ref{eq:2.15}).  This is a fully off-shell amplitude depending in
addition on the Poincar\'e-Jacobi momentum ${\bf q}$ of the pair.  The
embedded
2N transition amplitude satisfies the Lippmann-Schwinger equation
\begin{equation}
T_1({\bf k},{\bf k}';{\bf q}) = V_1({\bf k},{\bf k}';{\bf q}) + \int d
{\bf k}'' \: \frac{V_1({\bf k},{\bf k}'';{\bf q}) \: T_1({\bf k}'',{\bf
k}';{\bf q})} {\sqrt{[2\sqrt{m^2+ \mathbf{p}'^2}]^2 + \mathbf{q}^2} -
\sqrt{[2\sqrt{m^2+\mathbf{p}''^2}]^2 +\mathbf{q}^2} +i\epsilon} .
\label{eq:3.1}
\end{equation}
where the interaction (\ref{eq:b.49}) can be expressed in
the relativistic Jacobi momenta as Ref.~\cite{fch}
\begin{equation}
V({\bf q}) = \sqrt{\left[2\sqrt{m^2+\mathbf{k}^2} + v\right]^2 +
\mathbf{q}^2} \: - \: \sqrt{\left[2\sqrt{m^2+\mathbf{k}^2}\right]^2
+\mathbf{q}^2}.
\label{eq:3.2}
\end{equation}
For ${\bf q}=0$ this expression  reduces to the interaction 
$v({\bf k},{\bf k}')\delta (\mathbf{q}-\mathbf{q}\,')$, which is the 
interaction in the two-nucleon mass operator.
In the same limit,  Eq.~(\ref{eq:3.1}) reduces to the familiar
Lippmann-Schwinger equation with relativistic kinetic energies.

This matrix element is constructed using the methods outlined in
equations (\ref{eq:c.8}) and (\ref{eq:c.10}).  First the matrix element
of the right half-shell embedded $t$-operator is evaluated using the
two-body half shell transition amplitude where the convention of
\cite{bh} is employed
\begin{eqnarray}
 \langle \mathbf{k} | T_1(\mathbf{q};z') | \mathbf{k}' \rangle
&=& \langle \mathbf{k} | V(\mathbf{q}) | \mathbf{k}'^{(-)} \rangle
\nonumber \\
&=& \frac{2(E_{k'} + E_{k})}
{\sqrt{4E_{k'}^2 + \mathbf{q}^2}
    +\sqrt{4E_{k}^2 + \mathbf{q}^2}}
t(\mathbf{k},\mathbf{k}'; 2E_{k'}) ,
\label{eq:3.5}
\end{eqnarray}
where the 2N transition amplitude
$t(\mathbf{k},\mathbf{k}'; 2E_{k'})$ is the solution of the half-shell
Lippmann-Schwinger equation
\begin{eqnarray}
t(\mathbf{k},\mathbf{k}';2E_{k'})
= v( \mathbf{k}, \mathbf{k}')  + \int d\mathbf{k}''
 \frac{ v( \mathbf{k}, \mathbf{k}'') t(\mathbf{k}'',\mathbf{k}';2E_{k'}) 
}
{E_{k'} - 2\sqrt{m^2 +k''^2} + i\epsilon}.
\label{eq:3.6}
\end{eqnarray}
This solution is used as input to equation (\ref{eq:c.10}) which 
has the form
\begin{eqnarray}
 \langle \mathbf{k} |T_1(\mathbf{q};z)| \mathbf{k}' \rangle  &=&
 \langle \mathbf{k} | T_1(\mathbf{q};z'| \mathbf{k}' \rangle \; - \nonumber \\
 \int & d\mathbf{k}''& \langle \mathbf{k} | T_1(\mathbf{q};z)|
\mathbf{k}'' \rangle   \Big( \frac{1}{z - \sqrt{4(m^2 + \mathbf{k}''^2)
+ \mathbf{q}^2} }  - \frac{1}{z' - \sqrt{4(m^2 +
\mathbf{k}''^2) + \mathbf{q}^2} } \Big) \langle \mathbf{k}'' |
T_1(\mathbf{q};z') | \mathbf{k}' \rangle,
\label{eq:3.4}
\end{eqnarray}
where $T_1(z')$ is taken to be right half-shell with $z'=\sqrt{4(m^2 +
\mathbf{k}'^2) + \mathbf{q}^2} +i\epsilon$.  Note that in this 
equation the unknown matrix element is to the {\it left} of the kernel.
 
We refer to $T_1({\bf k},{\bf k}';{\bf q}):=
\langle \mathbf{k} |T_1(\mathbf{q};z)| \mathbf{k}' \rangle$ 
as the embedded 2N
t-matrix and to $V({\bf k},{\bf k}\,';{\bf q})$ as the embedded
interaction.  Matrix elements of $T_1(z)$ can be alternatively calculated by
inserting a complete set of eigenstates of the 2N mass operator
$m_{12} = 2\sqrt{\mathbf{k}^2 +m^2} +v$, as has been carried out in
Ref.~\cite{reltrit02} for a relativistic calculation of the triton
binding energy using two-body s-waves. Similarly, this method of
spectral decomposition can be used to directly calculate the matrix
elements of the embedded two-body t-matrix, as has been done in
another relativistic calculation of the triton binding energy
\cite{reltrit86}. The general difficulty with this method of spectral
decomposition is that it requires an integration over the
c.m. half-shell matrix elements $t({\bf k},{\bf k}'; E_{k'})$ in $k'$,
requiring the knowledge of those matrix elements for large values of
$k'$, which can pose a challenge with respect to numerical accuracy.
To our knowledge, this method has not yet led to a successful
relativistic calculation of scattering observables.

We use Eq.(\ref{eq:3.4}) to explicitly construct the elements of the
fully off-shell t-matrix, which enters the calculation of the
three-body transition amplitude given in Eq.~(\ref{eq:2.15}). For every
off-shell momentum $\mathbf{k}'$ the integral equation,
Eq.(\ref{eq:3.4}), must be solved for each $z$. It is worthwhile to
note that the $\mathbf{k}''$ integration in Eq.(\ref{eq:3.4}) only
involves momenta of the half-shell t-matrices, but {\it no}
energies. The momenta $\mathbf{k}$ and $\mathbf{k}'$ are fixed by
requirements of the three-body calculation, and typically are not
higher than 7~GeV. We tested that for converged results the
$\mathbf{k}''$ integration has to go up to about 12~GeV.  The
singularities in the two denominators of Eq.(\ref{eq:3.4}) do not pose
any problems and are handled with standard subtraction techniques.

In order to obtain insight into the impact of the embedding
for different values of $\mathbf{q}$, we introduce
approximations to the embedded interaction.
First, we completely neglect $\mathbf{q}$
in the embedded interaction, which leads to
\begin{equation}
V({\bf k},{\bf k}';{\bf q}) \to V_0({\bf k},{\bf k}';{\bf q}) =v({\bf
k},{\bf k}').
\label{eq:v0}
\end{equation}
Furthermore, we want to test the leading order terms in a $\mathbf{q}/m$
and $v/m$ expansion as
suggested in Ref.~\cite{witalar1}
\begin{equation}
V({\bf k},{\bf k}';{\bf q}) \to V_1({\bf k},{\bf k'};{\bf q}) =
v({\bf k},{\bf k}') \left(1-\frac{\mathbf{q}^2}{8m^2}\right)
\label{eq:v1}
\end{equation}
and
\begin{equation}
V({\bf k},{\bf k}';{\bf q}) \to V_2({\bf k},{\bf k}';{\bf q}) =
v({\bf k},{\bf k})
\left( 1- \frac{\mathbf{q}^2}{8\sqrt{m^2+\mathbf{k}^2} \:
\sqrt{m^2+\mathbf{k}'^2}} \right)
\label{eq:v2}
\end{equation}
and explore their validity as function of projectile energy.


\section{Cross Sections for Elastic Scattering}

The calculation of the cross section for elastic scattering,
Eq.~(\ref{eq:2.6}), requires the knowledge of the matrix element
$\langle \varphi_d, {\bf {\hat q}}, q_0|U|\varphi_d , {\bf q_0}\rangle$.
Using the definition of the operator $U$, Eq.~(\ref{eq:2.3}), inserting a complete
set of states and using for the matrix elements of the permutation
operator the expression from Eq.~(\ref{eq:2.11}), we obtain

\begin{eqnarray}
\langle \varphi_d, \mathbf{q} \Vert U \Vert \varphi_d, \mathbf{q}_0 \rangle &=&
\langle \varphi_d, \mathbf{q} \Vert P(z-M_0) + PT \Vert \varphi_d,
\mathbf{q}_0 \rangle \nonumber \\ &=& \int { d\mathbf{k}'
d\mathbf{q}'d\mathbf{k}''d\mathbf{q}''
\langle \varphi_d ,\mathbf{q} | \mathbf{k}', \mathbf{q}' \rangle \langle
\mathbf{k}', \mathbf{q}' \Vert P \Vert \mathbf{k}'', \mathbf{q}'' \rangle
\langle \mathbf{k}'', \mathbf{q}'' \Vert (z-M) + T \Vert \varphi_d ,
\mathbf{q}_0 \rangle } \nonumber \\ &=& \left(
W-\sqrt{m^2+\mathbf{q}_0^2} -\sqrt{4(m^2+(\mathbf{q} +
\frac{1}{2}\mathbf{q_0} \:
C(\mathbf{q},\mathbf{q}_0))^2)+\mathbf{q}_0^2} \right) \nonumber \\ &
& \times 2 N(\mathbf{q},\mathbf{q}_0) \varphi_d\left( \left|
\mathbf{q}_0 + \frac{1}{2}\mathbf{q} \:
C(\mathbf{q}_0,\mathbf{q})\right| \right) \varphi_d\left(\left|
\mathbf{q} + \frac{1}{2}\mathbf{q}_0 \:
C(\mathbf{q},\mathbf{q}_0)\right| \right) \nonumber \\ & & + 2 \int
d^3q'' \; N(\mathbf{q},\mathbf{q}'') \varphi_d\left( \left|
\mathbf{q}'' + \frac{1}{2}\mathbf{q} \:
C(\mathbf{q}'',\mathbf{q})\right| \right) \nonumber \\ & & \times \:
\langle \mathbf{q} + \frac{1}{2}\mathbf{q}'' \:
C(\mathbf{q},\mathbf{q}''),\mathbf{q}'' \Vert T \Vert \varphi_d, \mathbf{q}_0
\rangle .
\label{eq:4.1}
\end{eqnarray}
In first order the transition amplitudes reads $T=tP$, thus the final
expression for the transition amplitude for elastic scattering becomes
\begin{eqnarray}
\langle \varphi_d ,\mathbf{q} \Vert U \Vert \varphi_d ,\mathbf{q}_0 \rangle &=&
\left( {\sf W}-\sqrt{m^2+\mathbf{q}_0^2} -\sqrt{4[m^2+(\mathbf{q} +
\frac{1}{2}\mathbf{q}_0 \:
C(\mathbf{q},\mathbf{q}_0))^2]+\mathbf{q}_0^2} \right) \nonumber \\ &
& \times 2 N(\mathbf{q},\mathbf{q}_0) \varphi_d\left( \left|
\mathbf{q}_0 + \frac{1}{2}\mathbf{q} \:
C(\mathbf{q}_0,\mathbf{q})\right| \right) \varphi_d\left(\left|
\mathbf{q} + \frac{1}{2}\mathbf{q}_0 \:
C(\mathbf{q},\mathbf{q}_0)\right| \right) \nonumber \\ & & + 2 \int
d\mathbf{q}''  \; N(\mathbf{q},\mathbf{q}'') \varphi_d\left( \left|
\mathbf{q}'' + \frac{1}{2}\mathbf{q} \:
C(\mathbf{q}'',\mathbf{q})\right| \right) \: \varphi_d\left( \left|
\mathbf{q}'' + \frac{1}{2}\mathbf{q} \:
C(\mathbf{q}'',\mathbf{q}_0)\right| \right) \nonumber \\ & & \times
T_s \left( ({\bf q} + \frac{1}{2}\mathbf{q}'' \:
C(\mathbf{q},\mathbf{q}'')), ({\bf q}_0 + \frac{1}{2}\mathbf{q} \:
C(\mathbf{q}_0,\mathbf{q})); \varepsilon\right),
\label{eq:4.2}
\end{eqnarray}
where $\varepsilon = {\sf W} - \sqrt{m^2-\mathbf{q}''^2}$.

In the following we want to compare a non-relativistic first order
calculation to a corresponding relativistic one.  What is common in
both calculations is the input two-body interaction.  In the
relativistic case it is transformed to be two-body scattering
equivalent to the non-relativistic two-body calculation.  Though we
consider only spin-isospin independent interactions, we nevertheless can
provide qualitative insights for various cross sections in three-body
scattering in the intermediate energy regime, which we define as
ranging from 200~MeV to 1~GeV projectile energy. The focus of our
investigation will be how kinematic and dynamic relativistic effects
manifest themselves at different energies and for different scattering
observables.

As model interaction we choose a superposition of two Yukawa
interactions of Malfliet-Tjon type \cite{malfliet69-1} with parameters
chosen such that the potential supports a bound-state, the `deuteron',
at -2.23~MeV. The parameters are given in Ref.~\cite{scatter3d}.  With
this interaction we solve the non-relativistic Faddeev equation in first
order as a basis for all comparisons. Then we need to construct a
phase equivalent relativistic two-body interaction.  We
use the procedure suggested by Kamada-Gl\"ockle \cite{KG}, and obtain
a two-body interaction $v({\bf k},{\bf k}')$ as Born term of a
relativistic two-body Lippmann-Schwinger equation. This two-body t-matrix,
$t({\bf k},{\bf k}';\varepsilon)$ is the starting point for all calculations
which will be presented in the following. In principle there are other
methods to obtain a phase-shift equivalent relativistic potential
\cite{keipo}, however in this work we want to focus on the
relativistic effects visible in three-body scattering observables,
and thus use only one fixed scheme.

\begin{figure}
\begin{center}
\includegraphics[width=10cm]{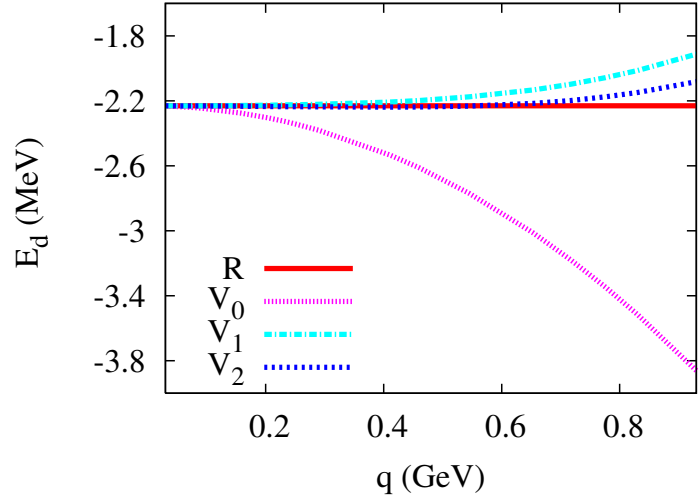}
\end{center}
\caption{(Color online) The deuteron binding in energy calculated with the embedded
interaction $V({\bf k},{\bf p}; {\bf q})$ as function of
{\bf q}. The solid line labeled `R' represents the binding energy of
-2.23~MeV which is independent of {\bf q}, when the full embedded interaction is
employed. The
dotted line is obtained if {\bf q} is set to zero in the embedded
interaction.  The dotted, dash-dotted and dashed lines show the approximations
to the embedded interaction $V_0$, $V_1$, and $V_2$ as given in
Eqs.~(\ref{eq:v0}), (\ref{eq:v1}), and (\ref{eq:v2}).
\label{fig3}
}
\end{figure}

Following Ref.~\cite{witalar1}, as first assessment of the quality
of different approximations for the embedded interaction we solve
the relativistic 2N Schr\"odinger equation for the deuteron as
function of the momentum $\vert{\bf q}\vert$, which takes the form

\begin{equation}
\Phi_d({\bf k}) = \frac{1}{\sqrt{m_d^2 +{\bf q}^2} - \sqrt{2E_{k_m}^2
 +{\bf q}^2}} \; \int d{\bf k}' \; V({\bf k},{\bf k}';{\bf q})
 \Phi_d({\bf k}').
\label{eq:Ed}
\end{equation}
Here $m_d$ is the rest mass of the deuteron.  In Fig.~\ref{fig3} we
show the deuteron binding energy $\varepsilon_d$ calculated using the
approximations of the embedded interaction given in
Eqs.~(\ref{eq:v0}), (\ref{eq:v1}), and (\ref{eq:v2}). A correctly
embedded interaction should of course not change $\varepsilon_d$ at
all. We see that $\varepsilon_d$ based on the calculation using $V_0$
starts to deviate already for very small $q$. The approximation $V_1$
gives reasonable results up to $q \approx$~0.3~GeV, whereas $V_2$ is
good to about 0.6~GeV.  In the following we will see how far these
simple estimates are reflected in the calculation of various
scattering observables.

As a first observable we consider the total cross section for elastic
scattering, $\sigma_{el}$, which is given in Table~\ref{table-1} for
projectile kinetic energies from 10~MeV up to 1~GeV. Starting from the
non-relativistic cross section, we successively implement different
relativistic features to study them in detail.  First we only change
the phase space factor in the calculation (psf) together with the
relativistic transformation from laboratory to c.m. frame, and only
then implement the relativistic kinematics due to the
Poincar\'e-Jacobi coordinates (R-kin). The relativistic phase space
factor alone has a large effect on the size of the total cross
section, as was already observed in~\cite{ndtotal}. The kinematic
effects of the Poincar\'e-Jacobi coordinates have the opposite effect
and lower the cross section. However, all kinematic effects taken
together increase the total cross section by about 6\% at 0.2~GeV and
about 40\% at 1~GeV.  Introducing relativistic dynamic effects into
the calculation changes this considerably . The full relativistic
calculation (R) lowers the total cross section by about 2\% at 0.2~GeV
and about 6\% at 1~GeV, so that in total the relativistic cross
section is smaller than the nonrelativistic one.  The approximation
$V_2$ of Eq.~(\ref{eq:v2}) is very good in the energy regime
considered, even at 1~GeV its result only deviates by about 2\% from
the full one. As suggested by the calculations of the deuteron binding
energy, the approximation $V_1$ of Eq.~(\ref{eq:v1}) is still
reasonable at 0.2~GeV, but after that starts to become worse.

\begin{figure}
\begin{center}
\includegraphics[width=10cm]{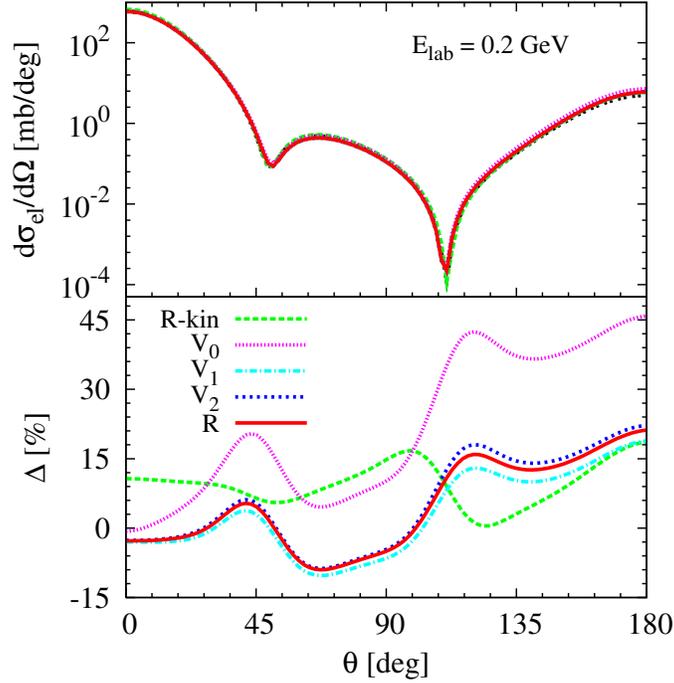}
\end{center}
\caption{(Color online) The differential cross section for elastic scattering at 0.2~GeV projectile
kinetic energy as function
of the laboratory scattering angle (upper panel). The solid line represents the fully
relativistic calculation.
The lower panel shows the relative deviation $\Delta$ with respect to the corresponding
non-relativistic calculation. The long-dashed curve labeled `R-kin' represents a
calculation in which  only relativistic kinematic effects are incorporated. The dotted,
dash-dotted, and  dashed curves show the approximations to the
embedded interaction $V_0$, $V_1$, and $V_2$ as given in
Eqs.~(\ref{eq:v0}), (\ref{eq:v1}), and (\ref{eq:v2}).
\label{fig4} }
\end{figure}

\begin{figure}
\begin{center}
\includegraphics[width=10cm]{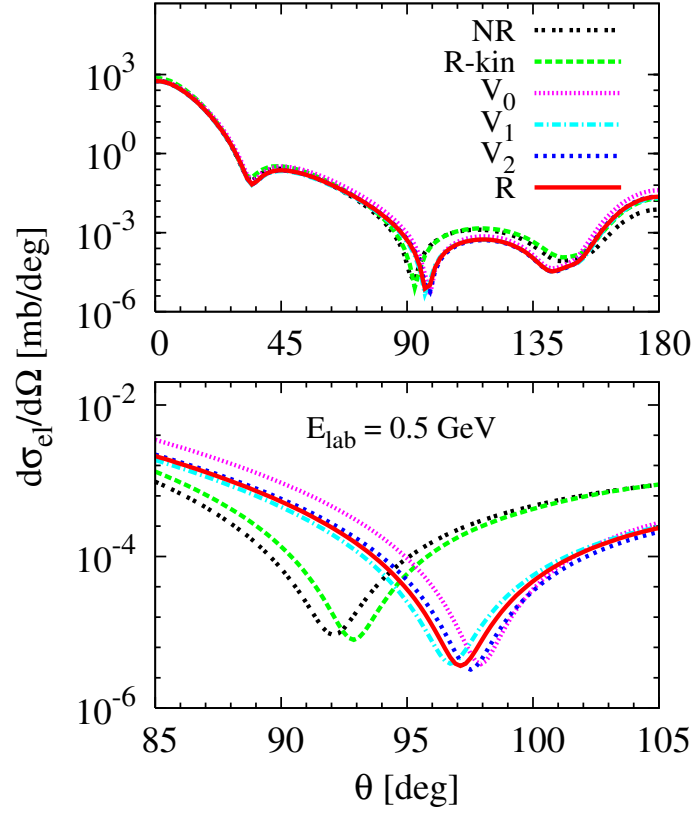}
\end{center}
\caption{(Color online) The differential cross section for elastic scattering at 0.5~GeV projectile
kinetic energy as function
of the laboratory scattering angle. The double-dotted curve labeled `NR' represents the
non-relativistic
calculation, and the solid curve labeled `R' the corresponding fully relativistic one.
The long-dashed curve labeled `R-kin' represents a
calculation in which  only relativistic kinematic effects are incorporated. The dotted,
dash-dotted, and  dashed curves show the approximations to the
embedded interaction $V_0$, $V_1$, and $V_2$ as given in
Eqs.~(\ref{eq:v0}), (\ref{eq:v1}), and (\ref{eq:v2}).
\label{fig5} }
\end{figure}

\begin{figure}
\begin{center}
\includegraphics[width=10cm]{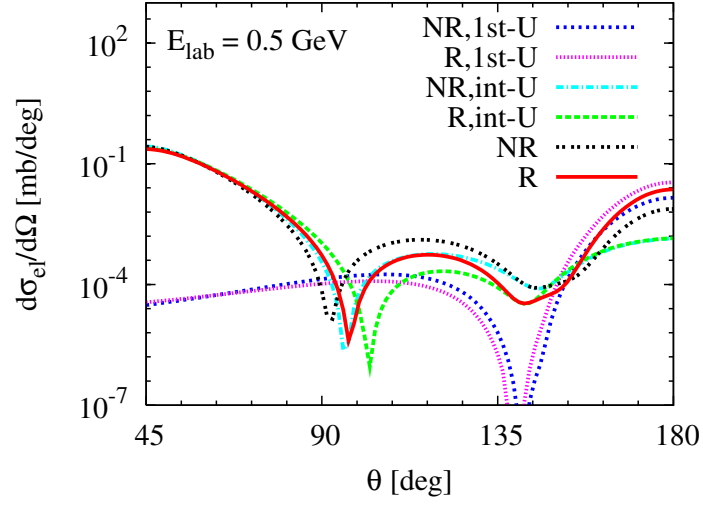}
\end{center}
\caption{(Color online) The differential cross section for elastic scattering at 0.5~GeV projectile
kinetic energy as function
of the laboratory scattering angle.
The double-dotted curve labeled `NR' represents the non-relativistic
calculation, and the solid curve labeled `R' the corresponding fully relativistic one.
The two other sets of curves show the contributions from the two different terms
contributing to
the transition operator $U$ in the non-relativistic (NR) and relativistic (R)
calculation. See text
for further discussion.
\label{fig6} }
\end{figure}

Next we consider the differential cross section for elastic scattering.
In Fig.~\ref{fig4} we show the calculation for 0.2~GeV projectile kinetic energy.
Since differences between the calculations disappear on a logarithmic
scale, we also show the quantity

\begin{equation}
\Delta = \frac{\left(\frac{d\sigma}{d\Omega} \right)_R -
\left(\frac{d\sigma}{d\Omega}\right)_{NR} }
{\left(\frac{d\sigma}{d\Omega}\right)_{NR}}
\label{eq:4.3}
\end{equation}
expressed in percentage for the different approximations in the lower
panel of Fig.~\ref{fig4}. For the backward angles, $\theta \ge 135^o$,
which correspond to higher momentum transfer, all relativistic effects
increase the cross section.  Here it can be clearly seen that indeed
$V_0$ is a bad approximation, whereas $V_1$ and $V_2$ are of about the
same quality. We also see that there is a small difference between the
calculation based on $V_2$ and the full result.  Similar findings,
however without the full calculation, were presented in
Ref.~\cite{witalar1}. When going to higher projectile kinetic
energies, we expect that the effects increase. This is indeed so, as
shown in Fig.~\ref{fig5} for the differential cross section at 0.5~GeV
projectile kinetic energy.  Here the second minimum in the cross
section around 90$^o$ shows a shift towards larger angles once
relativistic dynamics is included. This phenomenon has been seen and
studied in some electron-deuteron scattering~\cite{DPSW} calculations.
To study this shift in more detail we show in the lower panel of
Fig.~\ref{fig5} a restricted angular range. Here we can see that the
relativistic kinematics produces a shift of the minimum by a few
degrees. The magnification shows that the approximations of the
embedded interaction oscillate by a few degrees around the full
solution. At the extreme backward angles, the relativistic cross
section is larger than the non-relativistic one, as was the case at
0.2~GeV.  In order to illuminate the details of the two minima of the
cross section at 0.5~GeV even further, we show in Fig.~\ref{fig6} the
two terms contributing to the operator $U$ for elastic scattering
separately.  The curves labeled `{\it 1st-U}' correspond to the first
term in Eq.~(\ref{eq:4.2}) or the operator $P(z-M_0)$ in
Eq.~(\ref{eq:2.3}), which contributes to the structure of the cross
section at backward angles and depends only on the product of two
deuteron wave functions evaluated at shifted momenta.  Here the
relativistic calculation is pushed slightly towards smaller angles
indicating the effect of the functions $C({\bf q},{\bf q_0})$.  The
second term in Eq.~(\ref{eq:4.2}), represented by the curves labeled
`{\it int-U}', contains in integral over the two-body t-matrix and a
product of deuteron wave functions, and basically determines the
structure of the cross section for angles up to about 100$^o$.  Here
we see the shift of the minimum towards higher angles for the
relativistic calculation. The interference of both terms in the
calculation gives the final pattern as seen in Fig.~\ref{fig5}.


\section{Cross Sections for Breakup Processes}

The calculation of the breakup cross section, Eq.~(\ref{eq:2.2}), requires the
knowledge of the matrix element
$\langle {\bf k},{\bf q} \Vert U_0 \Vert \phi_d ,{\bf q_0}\rangle$.  Energy
conservation requires that in Eq.~(\ref{eq:2.2}) ${\sf W}_f ={\sf W}_i 
\equiv {\sf W}  =
\sqrt{4(m^2+k^2) +q^2} +\sqrt{m^2+q^2}$. This gives a relation between
the momenta {\bf k} and {\bf q}. In fact, for each given {\bf q} the magnitude
of {\bf k} is fixed as $k_a \equiv \vert {\mathbf k_{a}} \vert 
=\frac{1}{2}\sqrt{{\sf W}^2 -3m^2
-2{\sf W}\sqrt{m^2+ q^2}}$. This leads to
\begin{eqnarray}
\sigma_{br}^{cm} = \frac{ (2\pi)^4 }{3}
\frac{E(q_0)E_d(q_0)}{q_0 {\sf W}}
\; \int d\Omega_p d\Omega_q dq \; q^2 \frac{k_{a}}{4}
\sqrt{4(m^2+ k_{a}^2)+q^2} \;
\left| \langle {\mathbf p}, {\mathbf q} \Vert U_0 \Vert \varphi_d ,
{\mathbf q}_0 \rangle\right|^2
\label{eq:6.1}
\end{eqnarray}

We will consider here the cross sections for two different breakup
processes, the inclusive breakup, where only one of the outgoing
particles is detected, and the full or exclusive breakup.  In order to
obtain the differential cross section for inclusive breakup, one still
needs to integrate over the solid angle of the undetected particle to
arrive at invariant cross section:
\begin{equation}
\frac{d^3 \sigma_{br}^{cm}}{d\Omega _q dE_q}
= \frac{ (2\pi)^4 }{3} \frac{E(q_0)E_d(q_0)E(q) q}{q_0{\sf W}}
\; \int d\Omega_k \frac{k_{a}}{4}
\sqrt{4(m^2+k_{a}^2)+q^2} \;
\left|\langle \mathbf{k}, \mathbf{q} \Vert U_0 \Vert \varphi_d, \mathbf{q}_0
\rangle \right|^2 .
\label{eq:6.2}
\end{equation}
Here we changed from the variable $dq$ to the more utilized $dE_q$.
The five-fold differential cross section for exclusive scattering,
where both particles are detected, is given by
\begin{equation}
\frac{d^5 \sigma_{br}^{cm}}{d \Omega_k d\Omega _q dE_q}
= \frac{ (2\pi)^4 }{3} \frac{E(q_0) E_d(q_0)E(q) \; q}{q_0{\sf W}}
 \;  \frac{p_{a}}{4}
\sqrt{4(m^2+k_{a}^2)+q^2} \;
\left|\langle \mathbf{k}, \mathbf{q} \Vert U_0 \Vert \varphi_d, \mathbf{q}_0
\rangle \right|^2 .
\label{eq:6.3}
\end{equation}

Next we need to explicitly evaluate the matrix element for breakup scattering, $\langle \mathbf{k}, \mathbf{q}
\Vert U_0 \Vert \varphi_d, \mathbf{q}_0 \rangle$,  with $U_0$ given in Eq.~(\ref{eq:2.3})
\begin{equation}
\langle \mathbf{k}, \mathbf{q} \Vert U_0 \Vert \varphi_d, \mathbf{q}_0 \rangle
=  \langle \mathbf{k}, \mathbf{q} \Vert T \Vert \varphi_d, \mathbf{q}_0 \rangle
+ \langle \mathbf{k}, \mathbf{q} \Vert P_{12}P_{23}T \Vert 
\varphi_d, \mathbf{q}_0 \rangle
+ \langle \mathbf{k}, \mathbf{q} \Vert P_{13}P_{23}T \Vert 
\varphi_d, \mathbf{q}_0 \rangle.
\label{eq:6.4}
\end{equation}
The two terms containing the permutations can be calculated
analytically, as we show explicitly for the second term using the
expressions of Eqs.~(\ref{eq:2.9}) and (\ref{eq:2.10}) for the
Poincar\'{e}-Jacobi coordinates

\begin{eqnarray}
\lefteqn{ \langle \mathbf{k}, \mathbf{q} \Vert P_{12}P_{23}T \Vert
\varphi_d, \mathbf{q}_0 \rangle } \nonumber \\ &=& \int d {\bf q_1} d
{\bf q} _2 d {\bf q}_3 \int d{\bf k'} d{\bf q'} \langle \mathbf{k},
\mathbf{q} \vert \mathbf{q}_1, \mathbf{q}_2, \mathbf{q}_3 \rangle
\langle \mathbf{q}_1, \mathbf{q}_2, \mathbf{q}_3 \Vert P_{12}P_{23}
\Vert \mathbf{k}', \mathbf{q}' \rangle \langle \mathbf{k}', \mathbf{q}'
\Vert T \Vert \varphi_d, \mathbf{q}_0 \rangle \nonumber \\ &=& \int d
{\bf q_1} d {\bf q} _2 d {\bf q}_3 \int d{\bf k'} d{\bf q'} \delta (
\mathbf{q}_1 + \mathbf{q}_2 + \mathbf{q}_3 -\mathbf{P} ) \; {\hat
n}({\bf q}_1;{\bf q}_2,{\bf q}_3) \; {\hat n}({\bf q}_2;{\bf q}_3,{\bf
q}_1) \nonumber \\ & & \times \delta( \mathbf{q} - \mathbf{q}_1)\;
\delta ( \mathbf{k} - \mathbf{k}_{23})\;
\delta( \mathbf{q}' - \mathbf{q}_2) \; \delta ( 
\mathbf{k}' - \mathbf{k}_{31}) \langle \mathbf{k}',
\mathbf{q}' \Vert T \Vert \varphi_d, \mathbf{q}_0 \rangle .
\label{eq:6.5}
\end{eqnarray}
Taking particle 1 as spectator we can evaluate the momenta ${\bf q}_i$
explicitly as
\begin{eqnarray}
\mathbf{q}_1 &=& \mathbf{q} \nonumber \\ \mathbf{q}_2 &=& \mathbf{k} -
\frac{\mathbf{q}}{2E(k)} \left( \frac{ -\mathbf{k} \cdot \mathbf{q}}
{\sqrt{(2E(k))^2+q^2} + 2E(k)} + E(k) \right) \nonumber \\
\mathbf{q}_3 &=& - \mathbf{k} - \frac{\mathbf{q}}{2E(k)} \left( \frac{
\mathbf{k} \cdot \mathbf{q}}{\sqrt{(2E(k))^2+q^2} + 2E(k) } + E(k)
\right) .
\label{eq:6.6}
\end{eqnarray}
From this ${\bf q}'$ and ${\bf k}'$ can be obtained as ${\bf q}'={\bf
q}_3$ and ${\bf k}'= {\bf k}_{31}$ by inserting the expressions of
Eq.~(\ref{eq:6.6}) into Eq.~(\ref{eq:2.9}) leading to
\begin{equation}
\langle \mathbf{k}, \mathbf{q} \Vert P_{12}P_{23}T \Vert \varphi_d ,
\mathbf{q}_0 \rangle = {\hat n}({\bf q}_1;{\bf q}_2 {\bf q}_3)\; {\hat
n}({\bf q}_2;{\bf q}_3 {\bf q}_1) \langle {\bf k}_{31}, {\bf q}_2 \Vert
T \Vert \varphi_d, \mathbf{q}_0 \rangle.
\end{equation}
In first order we have $T=tP$, and an explicit evaluation leads to
\begin{eqnarray}
\langle \mathbf{k}, \mathbf{q} \Vert& P_{12}P_{23}T& \Vert \varphi_d,
\mathbf{q}_0 \rangle  = \nonumber \\
& &  {\hat n}({\bf q}_1;{\bf q}_2 {\bf q}_3)\; {\hat
n}({\bf q}_2;{\bf q}_3 {\bf q}_1) \; N({\bf q}_2,{\bf q}_0) \; T_s
\left({\bf k}_{31}, {\bf q}_0 +\frac{C({\bf q}_0,{\bf q}_2)}{2}{\bf q}_2;
\varepsilon\right) \; \varphi_d \left({\bf q}_2 + \frac{C({\bf q}_2,{\bf
q}_0)}{2}{\bf q}_0 \right). 
\label{eq:6.7}
\end{eqnarray}
The functions $N({\bf q}_2,{\bf q}_0)$ and $C({\bf q}_0,{\bf q}_2)$
are defined in Eqs.~(\ref{eq:2.12}) and (\ref{eq:2.13}).  The last
term in Eq.~(\ref{eq:6.4}), $\langle \mathbf{k}, \mathbf{q} \Vert
P_{13}P_{23}T \Vert \varphi_d, \mathbf{q}_0 \rangle$, is calculated
analogously.  Having calculated the matrix element of $U_0$,
Eq.~(\ref{eq:6.4}), we can obtain the differential cross section for
inclusive as well as exclusive breakup scattering. The expressions for
the invariant cross sections in the laboratory variables are 
derived in Appendix
\ref{appendixb}.

\begin{figure}
\begin{center}
\includegraphics[width=15cm,height=15cm]{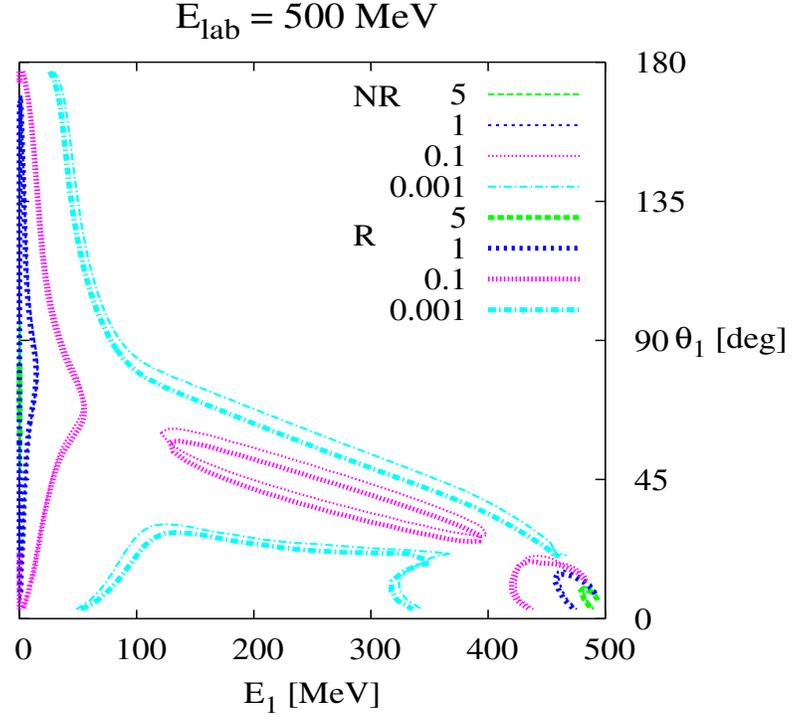}
\end{center}
\caption{
(Color online) The inclusive cross section at 0.5~GeV laboratory projectile kinetic energy
as function of the energy $E$ of the emitted particle and its emission angle
$\theta$.
The angles above the thin dashed line for larger energies and angles
are kinematically forbidden. The thin lines represent the contours of
the non-relativistic calculation, whereas the thick lines represent
the corresponding contours of the relativistic calculation.
\label{fig7} }
\end{figure}

\begin{figure}
\begin{center}
\includegraphics[width=14cm]{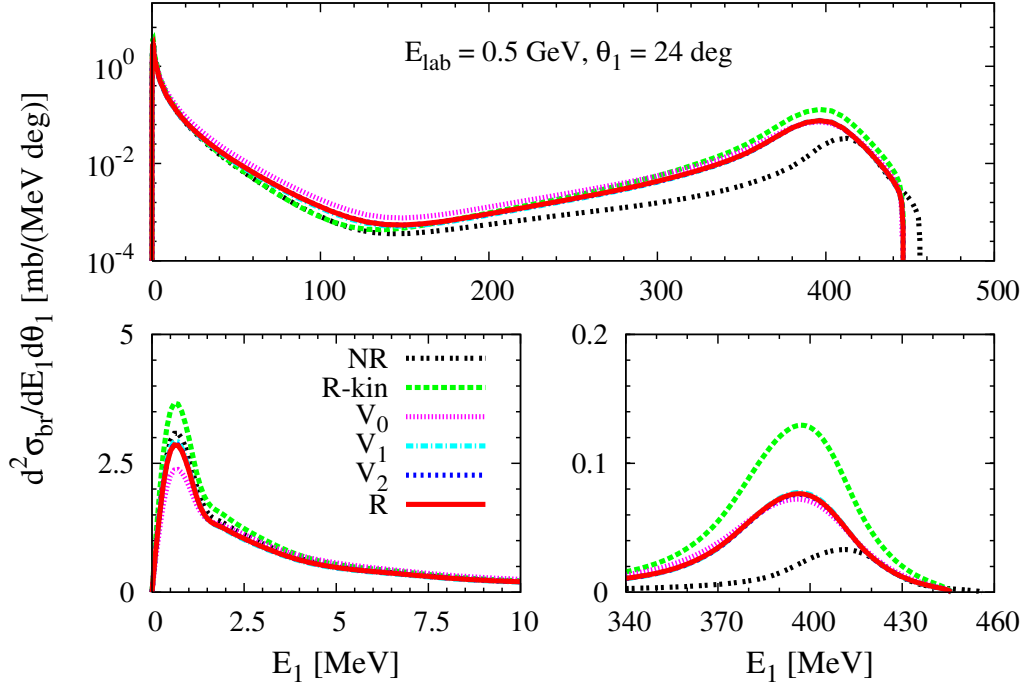}
\end{center}
\caption{
(Color online) The inclusive cross section at 0.5~GeV laboratory projectile kinetic energy
as function of the energy $E$ of the emitted particle and a 24$^o$ emission angle.
The upper panel displays the entire energy range of the emitted particle, whereas the
two lower panels show only the low and the high energies in a linear scale.
The solid line represents the fully relativistic calculation,
the double-dotted line the corresponding non-relativistic one. The long-dashed curve
labeled `R-kin' represents a
calculation in which  only relativistic kinematic effects are incorporated. The
dotted, dash-dotted,
 and  dashed curves show the approximations to the
embedded interaction $V_0$, $V_1$, and $V_2$ as given in
Eqs.~(\ref{eq:v0}), (\ref{eq:v1}), and (\ref{eq:v2}).
\label{fig8}
}
\end{figure}

\begin{figure}
\begin{center}
\includegraphics[width=10cm]{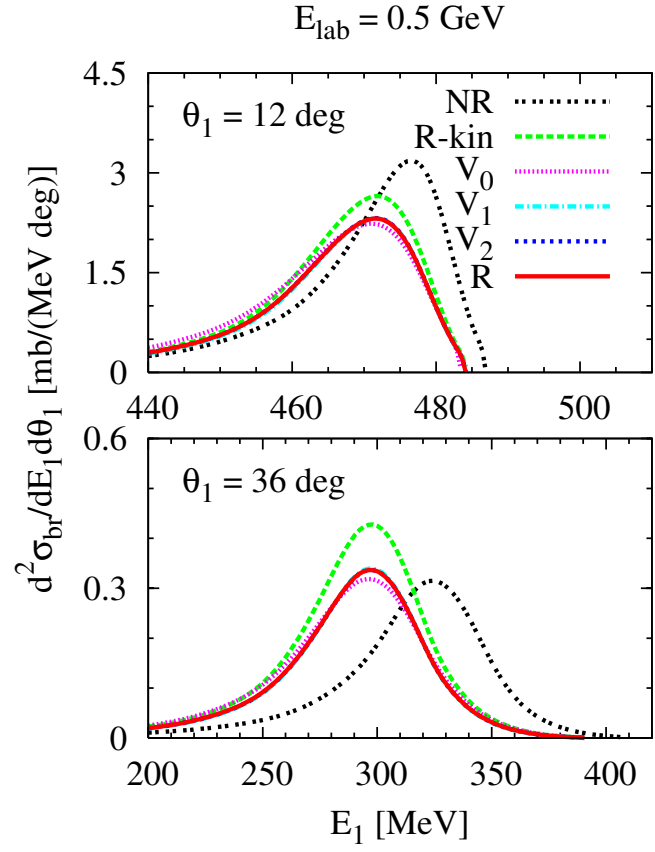}
\end{center}
\caption{
(Color online) Same as Fig.~\ref{fig8} but for different fixed angles of the emitted particle.
\label{fig9}
}
\end{figure}

\begin{figure}
\begin{center}
\includegraphics[width=10cm]{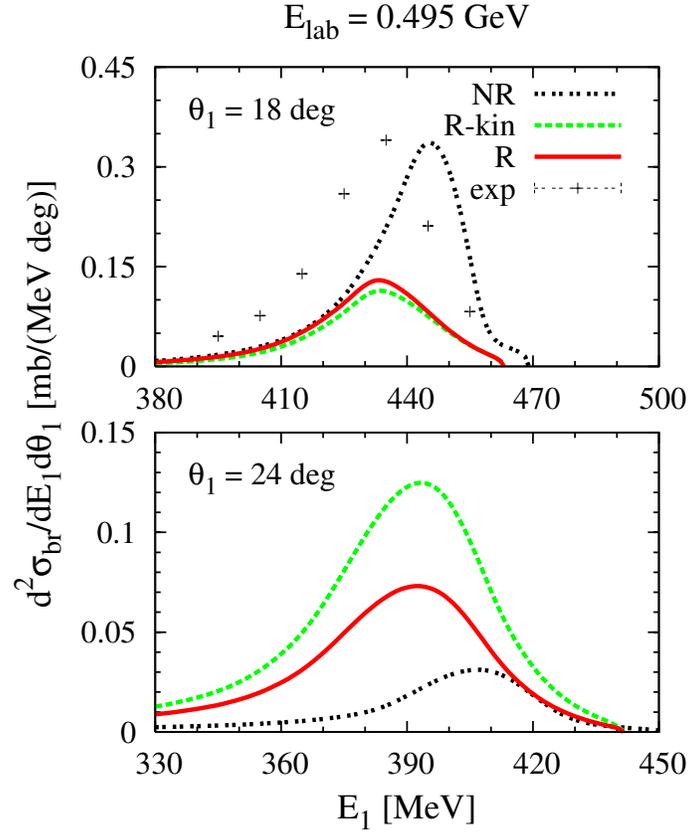}
\end{center}
\caption{
(Color online) The inclusive cross section at 0.495~GeV laboratory projectile
kinetic energy as function of the energy $E$ of the emitted particle
and fixed emission angles of 18$^o$ and 24$^o$ degrees. The solid line
represents the fully relativistic calculation and the double-dotted
line the non-relativistic one. The long-dashed curve labeled `R-kin'
represents a calculation in which only relativistic kinematic effects
are incorporated.  The data are from Ref.~\protect\cite{chen}.
\label{fig10}
}
\end{figure}

\begin{figure}
\begin{center}
\includegraphics[width=10cm]{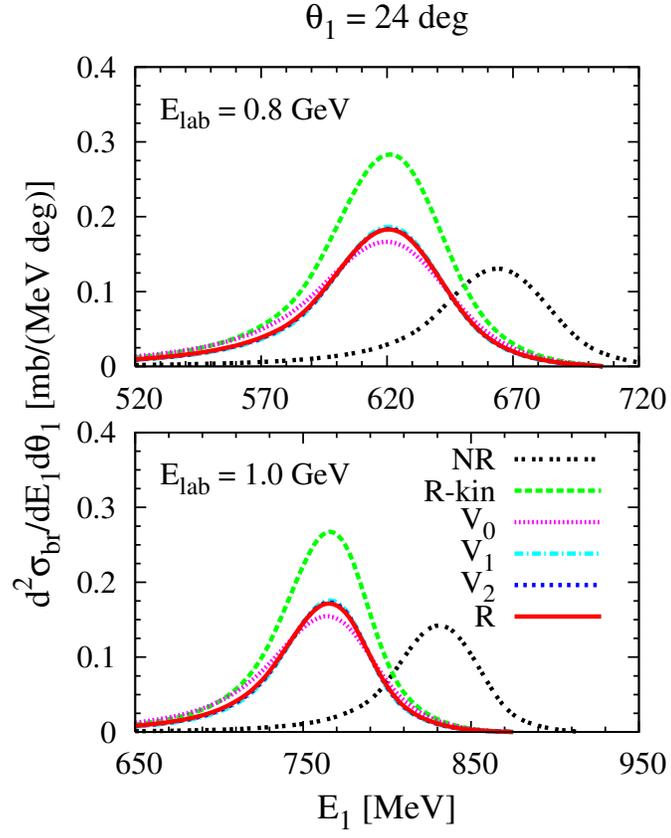}
\end{center}
\caption{(Color online)  The inclusive cross section at 0.8~GeV (upper panel)
and 1~GeV (lower panel) laboratory projectile kinetic energy as
function of the energy $E$ of the emitted particle and the fixed
emission angle of 24$^o$ degrees. The notation of the curves is the
same as in Fig.~\ref{fig9}.
\label{fig11}
}
\end{figure}

First, we consider inclusive breakup scattering and compare
the cross sections for a non-relativistic first-order calculation in
the two-body $t$-operator with the corresponding relativistic one. One
can expect that the evaluation of the delta function in the cross
section, Eq.~(\ref{eq:b.70}) will have a substantial effect on breakup
cross sections, since it fixes the relation between the magnitudes of
the vectors {\bf k} and {\bf q}. This in turn determines the maximum
allowed kinetic energy the ejected particle is allowed to have as
function of the emission angle. To get a global impression of those
differences Fig.~\ref{fig7} shows a contour plot of the differential
cross section for inclusive breakup scattering as function of the
kinetic energy and the emission angle of the ejected particle for the
non-relativistic and the fully relativistic calculation. The figure
shows that for each angle the maximum allowed kinetic energy of the
ejectile is shifted in the relativistic calculation towards smaller
values compared to the non-relativistic calculation. Specifically, one
can expect a shift of the quasi-free scattering (QFS) peak usually
studied in inclusive breakup scattering experiments.  In
Figs.~\ref{fig8} and \ref{fig9} we present specific cuts at different
constant angles to study details of the calculation. The upper panel
of Fig.~\ref{fig8} shows the entire energy range of the ejectile at
emission angle $\theta_1$~=~24$^o$ in a logarithmic scale, while the
lower two panels give a close-up of both peaks on a linear scale. The
QFS peak at the large ejectile energy exhibits clearly a shift towards
a slightly lower energy compared with the peak position calculated
non-relativistically. At this angle, relativistic kinematics given by
the phase-space factor and the Poincar\'{e}-Jacobi coordinates and
indicated by the line labeled `R-kin' results in a peak height, which
is almost double that of the full relativistic calculation shown as
solid line labeled `R'.  For breakup scattering we also study the
different approximations to the full calculation as introduced in
Section IV. In the QFS peak, which is defined by the condition that
one of the particles is at rest, even the crudest approximation $V_0$,
Eq.~(\ref{eq:v0}), is very close to the full calculation, the
approximations $V_1$ and $V_2$ are indistinguishable. This is not
surprising, since having one particle at rest means that the remaining
two are almost in their own c.m. frame, thus `boost' effects should be
extremely small.  Note that we work here in 
the total c.m. frame. It is quite illuminating to consider the second peak
at very small ejectile kinetic energies. Since the energy of the
particle is very small, it should become essentially non-relativistic.
This is indeed the case, and the full relativistic calculations almost
coincides with the non-relativistic one. It is worthwhile to note that
neither relativistic kinematics alone nor the approximation $V_0$,
which neglects the dependence of the embedded interaction on the
pair-momentum is close to the non-relativistic and full relativistic
calculations.  However, an approximate consideration of this
dependence as given by $V_1$ or $V_2$ seems to suffice.  We found that
this behavior is similar for low energy ejectiles, independent if the
energy of the projectile is 0.5~GeV or 1~GeV. In Fig.~\ref{fig9} we
show the QFS peak for two different angles in order to convey that the
increase or decrease of the height of the peak depends on the emission
angle under consideration.  In Fig.~\ref{fig10} we show the QFS peak
calculated for a projectile energy of 0.495~GeV and emission angles of
18$^o$ and 24$^o$, since there is experimental information available
for one of the angles. Here we see that the relativistic calculation
puts the peak at a position consistent with the data. Since we only
carry out a first order calculation with a model potential, we are not
surprised that the height of the peak is not described.  A similar
observation concerning the peak position was already made in
Ref.~\cite{Fachrud}, where a first order calculation with two
realistic NN interactions was carried out. To give an indication how
the position of the QFS peak shifts with increasing projectile energy,
we show in Fig.~\ref{fig11} the inclusive cross section for an
emission angle of 24$^o$ for projectile energies 0.8~GeV and
1.0~GeV. We see again that in the QFS regime the approximations $V_1$
and $V_2$ are essentially indistinguishable from the full
calculation. Considering only effects of relativistic kinematics
results in a peak height double as large as the full calculation,
indicating that dynamic effects are very important at those high
projectile energies.
  
\begin{figure}
\begin{center}
\includegraphics[width=13cm]{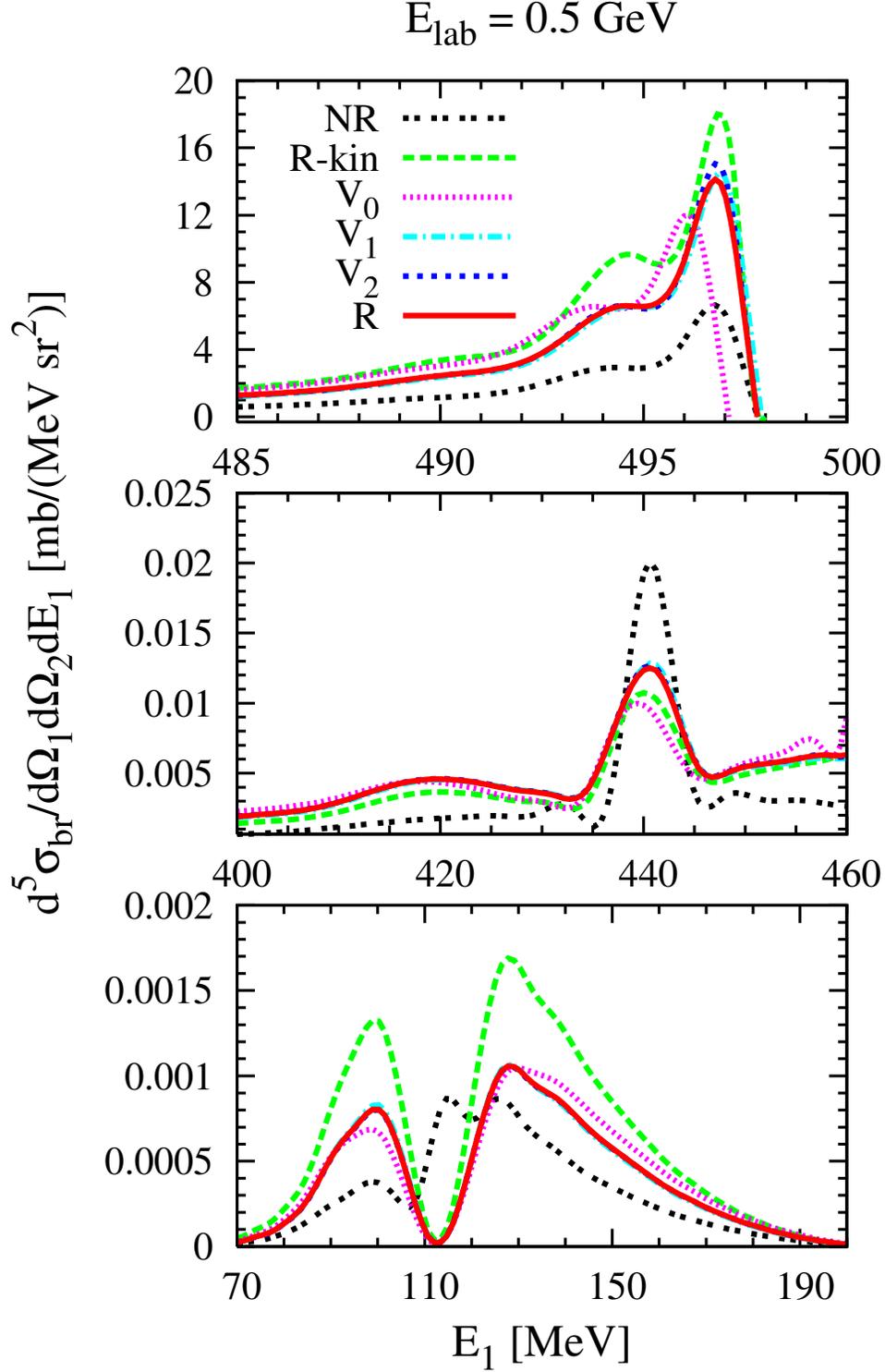}
\end{center}
\caption{(Color online) The exclusive breakup cross section at 0.5~GeV projectile kinetic energy
 as function of the ejected particle kinetic energy for
three different configurations defined in the three-body c.m. frame. For all configurations the angle
$\phi_{pq}$ is 0$^o$. For the upper panel the cos of the angle between {\bf q} and the beam ${\bf q_0}$ is 
$x_q=1$, i.e. the scattering occurs along the beam line, in the 
middle panel $x_q=\sqrt{3}/2$ and in the lower panel $x_q=-0.25$. The cos of the angle between {\bf p} and
${\bf q_0}$ is in the upper panel $x_p=0$, in the middle panel $x_p=-0.5$ and in the lower panel $x_p=-0.9$.
\label{fig12}
}
\end{figure}

\begin{figure}
\begin{center}
\includegraphics[width=13cm]{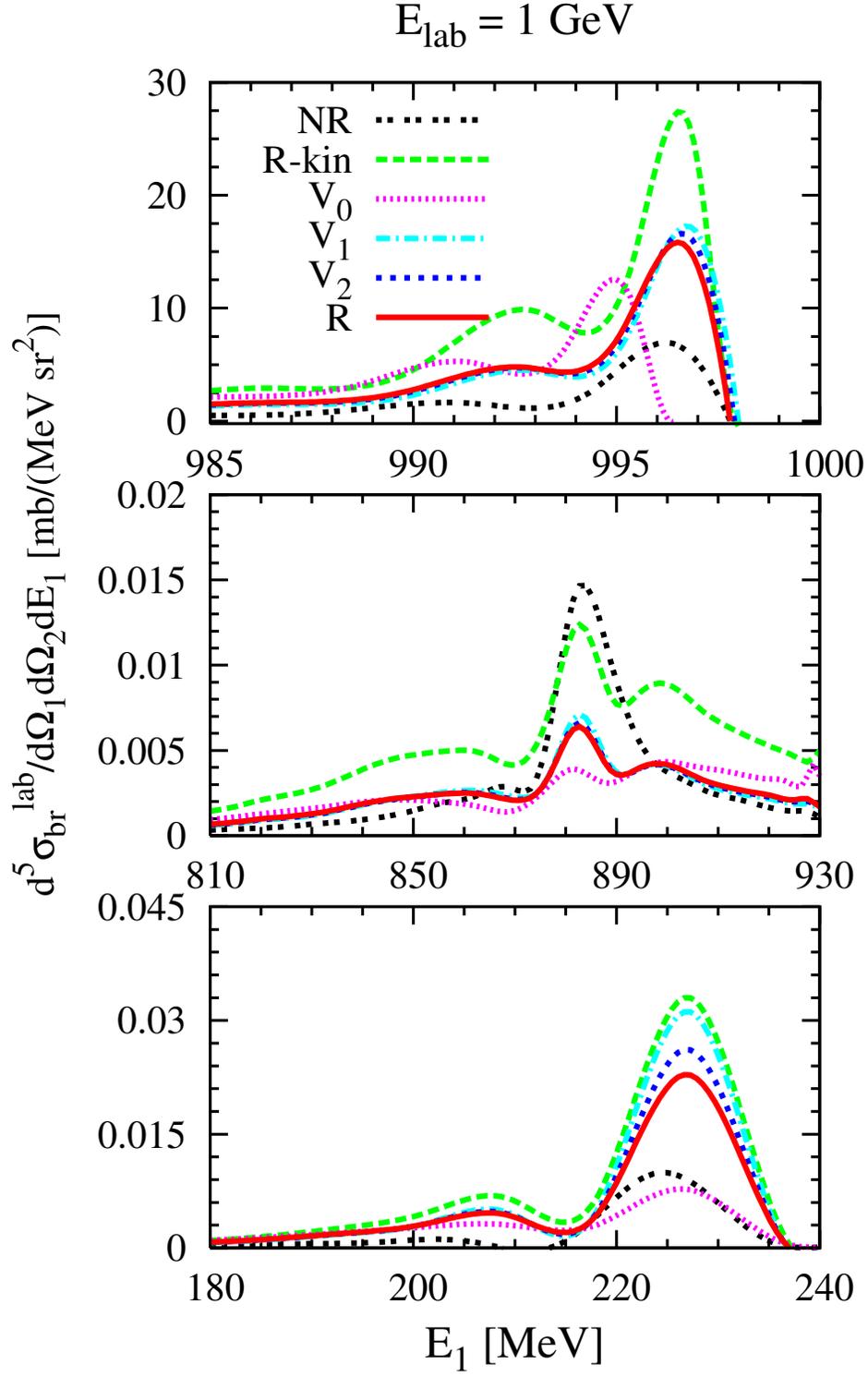}
\end{center}
\caption{(Color online) Same as Fig.\ref{fig12} but for projectile energy 1.0~GeV. 
\label{fig13}
}
\end{figure}

When considering exclusive breakup one faces many possible
configurations that could be considered. Since we are carrying out a
model study, we only want to show three specific
configurations at two selected energies, 0.5~GeV and 1~GeV in
Figs.~\ref{fig12} and \ref{fig13}. In all three configurations the
angle $\phi_{kq}$ between the projections of the vectors {\bf q} and
{\bf k} into the plane perpendicular to the beam direction ${\bf q_0}$
is zero.  Naive expectation is that when considering scattering in
first order in the $t$-operator the probability that one of the
particles is scattered along the beam is large. That is indeed the case as
shown in the upper panels of Figs.~\ref{fig12} and \ref{fig13},
depicting a so called collinear configuration in which the angle between
the vector {\bf q} and the beam direction ${\bf q_0}$ is zero (${\hat
{\bf q}}\cdot {\bf{\hat q}_0}\equiv x_q=1$) and the angle between the
vector {\bf p} and the beam direction is 90$^o$ (${\hat{\bf
k}}\cdot{\bf{\hat q}_0}\equiv x_k=0$).  Once the collinear condition
is no longer fulfilled, the cross section becomes considerably smaller, as
can be seen in the middle and lower panels of Figs.~\ref{fig12} and
\ref{fig13}. For the middle panels the angles are given by
$x_q=\sqrt{3}/2$ and $x_p=-0.5$, for the lower panels they are
$x_q=-0.25$ and $x_k=-0.9$. All configurations in Figs.~\ref{fig12}
and \ref{fig13} show considerable difference between the
nonrelativistic and relativistic calculations.  At 1~GeV we were
specifically looking for possible configurations where the
approximations $V_1$ and $V_2$ of Eqs.~(\ref{eq:v1}) and (\ref{eq:v2})
are not close to the full relativistic calculation any
longer. Considering the two-body binding energy displayed in
Fig.~\ref{fig3}, on should expect that the approximation $V_2$ can
exhibit deviations from the full result at 1~GeV.  One such
configuration is shown in the lower panel of Fig.~\ref{fig13}, where
there is a big difference between the calculation with $V_1$ and the
full result and a discernible difference between the calculation based
on $V_2$ and the full result.  However, we note, that in the majority
of configurations where the cross section is still relatively large,
$V_2$ is still a very good approximation at 1~GeV. Looking at the same
configuration at 0.5~GeV, $V_2$, and even $V_1$ are extremely good
approximations to the full result.


\section{Summary and Conclusions}

We investigated relativistic three-nucleon scattering 
with spinless interactions in the framework of
Poincar\'{e} invariant quantum mechanics. Since that framework is not
widely used in the nuclear physics few-body community we thought it adequate to
discuss the formulation  of that scheme in some detail, as well as the
formulation of scattering theory in this framework. The main points
are the construction of unitary irreducible representations of the
Poincar\'{e} group, both for noninteracting particles as well as for
interacting ones. The Poincar\'{e} interacting dynamics is constructed by
adding an interaction to the noninteracting mass operator which
commutes with and is independent of the total linear momentum and the
z-component of the total spin.

In this work we do not use partial waves but rather internal vector variables.
This leads to what we call Poincar\'{e}-Jacobi momenta. They
Wigner rotate under kinematic Lorentz transformations of the
underlying single particle momenta.
In the interacting three-body mass operator the two-body interactions
are embedded in the three-particle Hilbert space and are given as
\cite{fch}
\begin{equation}
V_{ij} := M_{ij} -M_0
= \sqrt{(m_{0,ij}+v_{ij})^2 + \mathbf{q}_{k}^2 }
- \sqrt{m_{0,ij}^2 + \mathbf{q}_{k}^2 }.
\end{equation}
This expression shows explicitly the dependence on the total momentum of the two-body system.
For the sake of completeness  we also discuss the multichannel scattering
theory in that relativistic framework and established the manifestly
invariant form of the differential cross section. 

The application to three-body scattering is based on the Faddeev
scheme, which is reformulated relativistically working with various
types of mass operators. The usage of the Poincar\'{e}-Jacobi momenta leads
to algebraic modifications of corresponding standard
nonrelativistic expressions, like e.g. the momentum representation of the
permutation operator,  Jacobians for the transitions between individual
 Jacobi momenta.  Due to the dependence on the total momentum of the
two-body interaction embedded in the three-body system, the two-body off-shell
t-operator entering the Faddeev equation acquires  additional
momentum dependence  beyond the usual energy shift which is characteristic in 
nonrelativistic calculations. This two-body t-operator is then evaluated by
expressing it exactly in terms of the solution of half-shell Lippmann-Schwinger
equations for a given two-body force in its c.m. frame.  
We also solve the relativistic  Lippmann-Schwinger equation for three
different 
momentum dependent two-body forces, which are approximations to the 
relativistic embedded interaction.

In order to compare a nonrelativistic calculation to a
relativistic one,  scattering  equivalent two-body forces in
the relativistic  and nonrelativistic formulations
have to be used. In this work we follow the KG prescription to arrive at
scattering equivalent two-body forces. There are different schemes,
and a detailed study on differences between those schemes will 
be subject of a forthcoming work.  We also 
restrict ourselves to a first order calculation in
the two-body t-operator, which however, already exhibits  most of the new
relativistic ingredients , both kinematically and dynamically. The
two-body force was chosen as a superposition  of two Yukawas of
Malfliet-Tjon type supporting a bound state, the deuteron, at
-2.23~MeV.  We calculate three-body scattering observables in the
intermediate energy regime, which we take to range from 0.2~GeV to
1~GeV. Those observables are cross sections for elastic as well as
breakup scattering, namely inclusive and exclusive scattering.  Not
surprisingly we find that the difference between nonrelativistic and
relativistic calculations increase with increasing energy. This is
specifically apparent when looking at the positions of minima in the
differential cross section as well positions of QFS peaks in inclusive
scattering.  When studying various approximations to the relativistic
embedded interaction, we find that if the approximations contain
terms proportional to the first order in a $p/m$ and $q/m$ expansion,
the approximation captures the features of the exact relativistic
calculation very well. Only at 1~GeV we start to find discernible
deviations in selected configurations for exclusive scattering.
Our results clearly indicate an interesting interplay of kinematically
and dynamically relativistic effects, which as expected, increase with
energy. For example, the total cross section is increased by kinematical 
effects, whereas the dynamical effects resulting from the q-dependence of the embedded
two-body force decrease it. This tendency is also seen in the height of the QFS peak
where in most cases the full calculation is lower than the one allowing only
for relativistic kinematic effects. Thus, considering only relativistic
kinematic effects leads in general to an over prediction of cross sections,
which becomes more dramatic the higher the projectile energy is.

The Poincar\'e invariant relativistic framework is formally close
to the nonrelativistic one and therefore standard nonrelativistic
formulations, in our case the Faddeev scheme, can be used with proper
modifications. The present restriction to a first order calculation in
the t-operator will soon be replaced by a complete solution of the
corresponding relativistic Faddeev equation following \cite{scatter3d}, where the
nonrelativistic Faddeev equation has been  successfully solved employing vector variables.
The application to the realistic world of pd scattering at the energies
up to 1 GeV considered in this study requires of course two-and
three-nucleon forces high above the pion production threshold. Though 
they are not
yet available,  they can also be included in the framework of 
Poincar\'e invariant quantum mechanics.


\vfill

\section*{Acknowledgments}
This work was performed in part under the
auspices of the U.~S.  Department of Energy under contract
No. DE-FG02-93ER40756 with Ohio University and
under contract
No. DE-FG02-86ER40286 with the University of Iowa. We thank
the Ohio  Supercomputer Center (OSC) for the use of
their facilities under grant PHS206.
The author's would like to acknowledge discussions with F. Coester that
contributed materially to this work.


\newpage
\appendix

\section{Invariance of $S$ and relation to $T$}
\label{appendixa}

The expression (\ref{eq:b.70}) and (\ref{eq:b.71}) for the differential cross section
can be rewritten in a manifestly invariant form.  We write them as a
product of an invariant phase space factor, an invariant factor
that includes the relative speed, and an invariant scattering amplitude.

To identify and establish the invariance of the invariant scattering
amplitude note that the scattering operator $S$ is Poincar\'e invariant:
\begin{eqnarray}
U_0 (\Lambda ,a) {\hat S} &=& U_0 (\Lambda ,a) \Omega_+^{\dagger} (H,H_0)
\Omega_- (H,H_0) = \Omega_+^{\dagger} (H,H_0) U (\Lambda ,a)
\Omega_- (H,H_0) \nonumber \\
&=&\Omega_+^{\dagger} (H,H_0) \Omega_- (H,H_0) U_0 (\Lambda ,a) =
{\hat S} U_0 (\Lambda ,a).
\label{eq:a.1}
\end{eqnarray}

The Poincar\'e invariance of the $S$ operator above is a
consequence of the intertwining relations for the wave operators
\begin{equation}
U (\Lambda ,a)
\Omega_\pm (H,H_0) =
\Omega_\pm (H,H_0) U_0 (\Lambda ,a)
\label{eq:a.2}
\end{equation}
To show the intertwining property of the wave operators first note
that the invariance principle gives the identity
\begin{equation}
\Omega_\pm (H,H_0) = \Omega_\pm (M,M_0) .
\label{eq:a.3}
\end{equation}
The mass operator intertwines by the standard intertwining properties
of wave operators.  For our choice of irreducible basis the
intertwining of the full Poincar\'e group follows because all of
the generators can be expressed as functions of the mass
operator and a common set of kinematic operators, $\{ \mathbf{K}, j_z
, j_x, j^2, -i\mbox{\boldmath$\nabla$}_K\}$, that commute with the
wave operators.

The covariance of the $S$ matrix elements follows from the Poincar\'e
invariance of the $S$ operator if the matrix elements of $S$ are
computed in a basis with a covariant normalization.

The $S$-matrix elements can be evaluated in the channel mass eigenstates.
After some algebra one obtains:
\begin{eqnarray}
\langle \beta \vert S_{ba} \vert \alpha \rangle 
&=&  \lim_{\tau \to \infty} \langle \beta \vert
e^{i M_\beta \tau} e^{-2iM\tau}e^{iM_\alpha \tau} \vert \alpha\rangle
\nonumber \\
&= &
\langle \beta \vert \alpha \rangle + \lim_{\tau \to \infty}
\int_0^\tau   d\tau'\, {d \over d\tau'}
\langle \beta \vert e^{i({\sf W}_\beta + {\sf W}_\alpha -2M )\tau'}
\vert \alpha \rangle
\nonumber \\
&= &
\langle \beta \vert \alpha \rangle \lim_{\epsilon  \to 0^+} \left
[ { 2i\epsilon \over {\sf W}_\beta - {\sf W}_\alpha + 2i\epsilon } \right ]
\nonumber \\
 & &+
\lim_{\epsilon  \to 0^+} \left  [ {-4i\epsilon \over ({\sf W}_\beta -
{\sf W}_\alpha)^2 + 4\epsilon^2} \right ] \langle \beta \vert
\left( V^\alpha + V^\beta G(\bar{\sf W}+ i \epsilon )V^\alpha
\right )\vert \alpha\rangle .
\label{eq:a.9}
\end{eqnarray}
where $M_\alpha \vert \alpha \rangle = {\sf W}_{\alpha} \vert \alpha
\rangle$ and $M_\beta \vert \beta \rangle = {\sf W}_{\beta} \vert
\alpha \rangle$ and $\bar{\sf W} := {1 \over 2} ({\sf W}_\alpha+{\sf
M}_\beta)$ is the average invariant mass eigenvalue of the initial
and final asymptotic states.  In deriving (\ref{eq:a.9}) the two
strong limits in (\ref{eq:b.55}) are replaced a single weak limit.
Equation (\ref{eq:a.9}) is interpreted as the kernel of an integral
operator.  $S$-matrix elements are obtained by integrating the sharp
eigenstates in Eq.~(\ref{eq:a.9}) over normalizable functions of the
energy and other continuous variables.  To simply this expression
define the residual interactions $V^\alpha$ and $V^\beta$ by:
\begin{equation}
V^\alpha := M - M_\alpha; \qquad V^\beta \:= M - M_\beta ,
\label{eq:a.5}
\end{equation}
where
\begin{equation}
V^\alpha \vert \alpha \rangle = (M-{\sf W}_\alpha)
\vert \alpha \rangle; \qquad
V^\beta \vert
\beta \rangle =
(M-{\sf W}_\beta) \vert \beta \rangle.
\label{eq:a.6}
\end{equation}
The resolvent operators of the mass operator and the
channel mass operator,
\begin{equation}
G(z) := {1 \over z-M} \qquad G_\alpha(z) := {1 \over z- M_\alpha},
\label{eq:a.7}
\end{equation}
are related by the second resolvent relations \cite{Hi}:
\begin{equation}
G(z) - G_\alpha(z) = G_\alpha(z)V^\alpha G(z) = G(z)V^\alpha G_\alpha(z).
\label{eq:a.8}
\end{equation}

It is now possible to evaluate the limit as $\epsilon \to  0$.
It is important to remember that this is the kernel of an integral
operator.

The first term in square brackets is unity when the initial and final
mass eigenvalues are identical, and zero otherwise; however, the limit
in the bracket is a Kronecker delta and {\it not} a Dirac delta
function.  For $\alpha \not= \beta$, $\langle \beta ({\sf W}') \vert
\alpha ({\sf W} ) \rangle$ are Lebesgue measurable in ${\sf W}'$ for
fixed ${\sf W}$, so there is no contribution from the first term in
Eq. (\ref{eq:a.9}).  For the case that ${\sf W}_\alpha = {\sf
W}_\beta$, we have $\langle \beta ({\sf W}') \vert \alpha ({\sf W})
\rangle\propto \delta ({\sf W}'-{\sf W} )$.  The matrix element
vanishes by orthogonality unless ${\sf W}_\beta = {\sf W}_\alpha$, but
then the coefficient is unity.  Thus, the first term in (\ref{eq:a.9})
is $\langle \beta | \alpha \rangle$ if the initial and final channels
are the same, but zero otherwise.  The matrix elements also vanish by
orthogonality for two different channels governed by the same
asymptotic mass operator with the same invariant mass.  The first term
in (\ref{eq:a.9})  therefore includes a {\it channel} delta function.

For the second term, the
quantity in square brackets becomes $-2 \pi i \delta ({\sf W}_\beta -
{\sf W}_\alpha)$, which leads to the relation
\begin{equation}
\langle \beta \vert S \vert \alpha \rangle = \langle \alpha \vert \beta \rangle
- 2 \pi i \delta ({\sf W}_\beta -
{\sf W}_\alpha) \langle
\beta \vert T^{\beta \alpha}({\sf W}_\alpha+i0^+) \vert \alpha \rangle,
\label{eq:a.10}
\end{equation}
where
\begin{equation}
T^{\beta\alpha}(z) = V^\alpha + V^\beta G(z) V^\beta.
\label{eq:a.11}
\end{equation}
and $\langle \alpha \vert \beta \rangle $ is zero
if the initial and final channels are different and is the overlap of
the initial and final states if the initial and final channels are the
same.  Equation (\ref{eq:a.10}) is exactly eq. (\ref{eq:b.62}).

With our choice of irreducible basis
the residual interactions and the resolvent
commute with the total linear momentum operator, and if the sharp channel
states $\vert \alpha \rangle$ and $\vert \beta \rangle$ are simultaneous
eigenstates of the appropriate partition mass operator and the linear
momentum, then a three-momentum conserving delta function can be
factored out of the $T$-matrix element:
\begin{equation}
\langle \beta \vert T^{\beta\alpha}({\sf W}_\alpha+i0^+)
\vert \alpha \rangle =
\delta^3 ( \mathbf{P}_\beta - \mathbf{P}_\alpha )
\langle \beta \Vert  T^{\beta \alpha}({\sf W}_\alpha+i0^+)
\Vert  \alpha \rangle.
\label{eq:a.12}
\end{equation}
When combined with the three-momentum conserving delta function
the invariant mass delta function can be replaced an energy
conserving delta function
\begin{equation}
\delta ({\sf W}_\beta - {\sf W}_\alpha) =
\left| {d{\sf W} \over d{\sf E}}  \right| \; \delta ({\sf E}_\beta - {\sf E}_\alpha)  \qquad 
\left| {d{\sf W} \over d{\sf E}}  \right| = 
{{\sf W} \over {\sf E}}
.
\label{eq:a.13}
\end{equation}
The $S$-matrix elements
can be expressed in terms of the {\it reduced channel
transition operators} as follows:
\begin{equation}
\langle \beta \vert S \vert \alpha \rangle = \langle \alpha \vert \beta
\rangle
\delta_{\beta \alpha} - i (2 \pi) \delta^4 (P_\beta - P_\alpha)
{{\sf W}_{\alpha} \over {\sf E}_{\alpha}}
\langle \beta \Vert T^{\beta\alpha}({\sf W}_\alpha+i0^+
) \Vert  \alpha \rangle
\label{eq:a.14}
\end{equation}
In this expression the $S$ operator is invariant while the single
particle asymptotic states have a non-covariant normalization.

To extract the standard expression for the invariant amplitude the
single particle states are replaced by states with the covariant
normalization used in the particle data book \cite{pdg}:
\begin{equation}
\vert \mathbf{p}, \mu \rangle \longrightarrow  \vert p, \mu \rangle_{cov} =
\vert \mathbf{p},\mu \rangle \sqrt{2 E_{p_m}}(2 \pi)^{3/2} .
\label{eq:a.15}
\end{equation}
The resulting expression
\begin{equation}
- i (2 \pi) \delta^4 (P_\beta - P_\alpha)
{{\sf W}_{\alpha} \over {\sf E}_{\alpha}}{}
_{cov}\langle \beta \Vert T^{\beta\alpha}({\sf W}_\alpha+i0^+)
\Vert \alpha \rangle_{cov}
\label{eq:a.16}
\end{equation}
is invariant (up to spin transformation properties).  Since the four
dimensional delta function is invariant, the factor multiplying the
delta function is also invariant (up to spin transformation properties).
This means that
\begin{equation}
_{cov}\langle \alpha \Vert M^{\alpha \beta} \Vert \beta \rangle_{cov} :=
{1 \over (2 \pi)^3}
{{\sf W}_{\alpha} \over {\sf E}_{\alpha}}{}
_{cov}\langle \beta \Vert T^{\beta\alpha}({\sf W}_\alpha+i0^+) \Vert
\alpha \rangle_{cov}
\label{eq:a.17}
\end{equation}
is a Lorentz covariant amplitude.  The factor of $1/(2 \pi)^3$ is chosen
to agree with the normalization convention used in the particle
data book \cite{pdg}.

The differential cross
section becomes
\begin{eqnarray}
{d \sigma } & = &
{(2\pi)^4 \over
4  E_{m_t} (\mathbf{p}_t)
E_{m_b} (\mathbf{p}_b) v_{bt}} \;
\left|_{cov}\langle p_1, \cdots , p_n, \Vert M^{\alpha \beta} \Vert
\bar{p}_b,  \bar{p}_t  \rangle_{cov} \right| ^2
\nonumber \\
 &\times &
\delta^4  \left( \sum_i p_i  - \bar{p}_b -\bar{p}_t \right) \;
{d\mathbf{p}_1 \over  2 E_{m_1} (2 \pi)^3 }
\cdots
{d\mathbf{p}_n \over  2 E_{m_n} (2 \pi)^3 }.
\label{eq:a.18}
\end{eqnarray}
The identity
\begin{equation}
v_{bt}  ={\sqrt{ (p_t\cdot p_b)^2 -m_b^2 m_t^2} \over
E_{m_b} E_{m_t} }
\label{eq:a.19}
\end{equation}
can be used to get an invariant expression for the relative speed between the
projectile and target and
\begin{equation}
d\Phi_n ({p}_b+{k}_t; \mathbf{p}_1, \cdots , \mathbf{p}_n) =
  \delta^4 \left( \sum_i p_i - \bar{p}_b -\bar{p}_t \right) \;
{d\mathbf{k}_1 \over 2 E_{m_1} (2 \pi)^3 } \cdots
{d\mathbf{k}_n \over 2  E_{m_n} (2 \pi)^3 }
\label{eq:a.20}
\end{equation}
is the standard Lorentz invariant phase space factor.  Inserting these
covariant expressions in the definition of the differential cross
section gives the standard formula for the invariant cross section
\begin{equation}
d \sigma =
{(2 \pi)^4 \over
4 \sqrt{ (p_t\cdot p_b)^2 -m_b^2 m_t^2}}
\left|_{cov}\langle p_1, \cdots , p_n, \Vert M^{\alpha \beta} \Vert
\bar{p}_b,  \bar{p}_t  \rangle_{cov} \right| ^2
d\Phi_n ({p}_b+{p}_t; \mathbf{p}_1, \cdots , \mathbf{p}_n).
\label{eq:a.21}
\end{equation}
Because of the unitarity of the Wigner rotations and the covariance of
$\left|_{cov}\langle p_1, \cdots , p_n, \Vert M^{\alpha \beta} \Vert
\bar{p}_b, \bar{p}_t \rangle_{cov} \right| ^2$ this becomes an
invariant if the initial spins are averaged and the final spins are
summed. 

This manifestly invariant formula for the cross section is identical
to (\ref{eq:b.70}) and (\ref{eq:b.71}); 
in this form the invariant cross section can
be evaluated in any frame. The index $t$ refers to the target, which is
in our case the deuteron.


\section{Breakup Cross Section in the Laboratory Frame
variables}
\label{appendixb}

The total breakup cross section is Lorentz invariant.  The expression
for the differential cross sections (\ref{eq:b.70}) is given in terms
of single particle variables, while the solutions of the Faddeev
equations give transition matrix elements as functions of the
Poincar\'e-Jacobi momenta defined in Section II.  To compute the 
total cross section it is useful to work in a single 
representation.  Since the single particle momenta are directly related
to measured parameters of the differential cross section, we 
change variable in the transition amplitudes from Poincar\'e Jacobi
moment to single particle momenta.  In this section the single particle
variables are computed in the laboratory frame.

The relation between the product of 
single particle basis states and states expressed in 
terms of the Poincar\'e-Jacobi momenta are given by 
\begin{eqnarray}
\vert \mathbf{p}_{n}, \mathbf{p}_d \rangle
&=& \left| \frac{ \partial(\mathbf{q}_0, \mathbf{P}_i) }
    { \partial(\mathbf{p}_{n}, \mathbf{p}_d) } \right|^{1/2}
\vert \mathbf{q}_0, \mathbf{P}_i \rangle 
\nonumber \\
\vert \mathbf{p}_1, \mathbf{p}_2, \mathbf{p}_3 \rangle
&=& \left| \frac{ \partial(\mathbf{k},  \mathbf{p}_2 +  \mathbf{p}_3) }
    { \partial(\mathbf{p}_2, \mathbf{p}_3) } \right|^{1/2}
\left| \frac{ \partial(\mathbf{q}, \mathbf{P}_f) }
    { \partial(\mathbf{p}_1,  \mathbf{p}_2 +  \mathbf{p}_3) } \right|^{1/2}
\vert \mathbf{P}_f , \mathbf{k}, \mathbf{q}  \rangle 
\label{eq:d.2}
\end{eqnarray}
The Jacobians in these transformations are
\begin{eqnarray}
\left | \frac{ \partial(\mathbf{q}_0, \mathbf{P}_i) }{ \partial(\mathbf{p}_{n}, \mathbf{p}_d) } \right |
&= &\frac {E(q_0)E_d(q_0) {\sf E}_i }{E(p_{n})E_d(p_d)W} 
\nonumber \\
\left | \frac{ \partial(\mathbf{k},  \mathbf{p}_2 +  
\mathbf{p}_3) }{ \partial(\mathbf{p}_2, \mathbf{p}_3) }\right |
&= & \frac{\sqrt{(E(k_2)+E(p_3))^2 -(\mathbf{p}_2 +  \mathbf{p}_3)^2} 
\;(E(p_2)+E(p_3))}
{4E(p_2)E(p_3)}
\nonumber \\
\left | \frac{ \partial(\mathbf{q}, \mathbf{P}_f) }{ \partial(\mathbf{p}_1,
\mathbf{p}_2 +  \mathbf{p}_3) }\right |
&= &\frac {\sqrt{(E(p_2)+E(p_3))^2 -(\mathbf{p}_2 +  \mathbf{p}_3)^2 + \mathbf{q}^2} \; E(q) {\sf E}_f }
{E(p_1) \; (E(p_2)+E(p_3)) \; W}
\label{eq:d.6}
\end{eqnarray}

\noindent Defining 
\begin{equation}
\Gamma (W,\mathbf{k},\mathbf{q}) = { W^2  \over {\sf E}_i^2}
\left| \frac{ \partial(\mathbf{q}_0, \mathbf{P}_i) }
    { \partial(\mathbf{p}_{n}, \mathbf{p}_d) } \right|
\left| \frac{ \partial(\mathbf{k},  \mathbf{p}_2 +  \mathbf{p}_3) }
    { \partial(\mathbf{p}_2, \mathbf{p}_3) } \right|
\left| \frac{ \partial(\mathbf{q}, \mathbf{P}_f) }
    { \partial(\mathbf{p}_1,  \mathbf{p}_2 +  \mathbf{p}_3) } \right| ,
\label{eq:d.7}
\end{equation}
the total cross section for breakup scattering becomes
\begin{eqnarray}
 \sigma_{br}^{lab} &=& \frac{1}{3} \frac{ (2\pi)^4 }{v_{bt}}
\int d\mathbf{p}_1d\mathbf{p}_2d\mathbf{p}_3   
 \; \delta({\sf E}_f-{\sf E}_i) \; \delta^3(\mathbf{P}_f-\mathbf{P}_i) 
\Gamma (W,\mathbf{p},\mathbf{q})
\left| \langle \mathbf{p}, \mathbf{q}  \Vert U_0 \Vert
 \varphi_d ,\mathbf{q}_0 \rangle \right| ^2 \nonumber \\
&=& \frac{1}{3} \frac{ (2\pi)^4 }{v_{bt}}
\int d\mathbf{p}_1d\mathbf{p}_2  
\;  \delta({\sf E}_f-{\sf E}_i)  \Gamma (W,\mathbf{p},\mathbf{q})
 \left| \langle \mathbf{p}, \mathbf{q}  \Vert U_0 \Vert
 \varphi_d, \mathbf{q}_0 \rangle  \right| ^2,
\label{eq:d.8}
\end{eqnarray}
where we used $\mathbf{P}=\mathbf{P}_i=\mathbf{P}_f$.
The $\delta$ function in the energy can be eliminated 
by a variable change.
The total energy of the system is ${\sf E}={\sf E}_i={\sf E}_f=
\sqrt{m^2 + \mathbf{p}_1^2} + \sqrt{m^2 + \mathbf{p}_2^2} + 
\sqrt{m^2 + (\mathbf{P}-\mathbf{p}_1-\mathbf{p}_2)^2}$.
With $ p_1 \equiv | {\bf p}_1 | $  and 
${\bf {\hat p}}_2 \equiv {\bf p}_2 / \vert  {\bf p}_2 \vert$ 
this gives 

\begin{equation}
\frac{d{\sf E}}{dp_2} = \frac{p_2}{E(p_2)}
+ \frac{p_2-(\mathbf{P}-\mathbf{p}_1) \cdot \hat {\mathbf{p}}_2}{{\sf
E}-E(p_1)-E(p_2)}
=\frac{p_2({\sf E}-E(p_1)) - E(p_2)(\mathbf{P}-\mathbf{p}_1) 
\cdot \hat {\mathbf{p}}_2 }{E(p_2)E(p_3)}.
\label{eq:d.9}
\end{equation}
Since $dp_1=E(p_1)/p_1 dE_1$, and $v_{bt}=p_{n}/E(p_{n})$,
Eq.~(\ref{eq:d.8}) becomes
\begin{equation}
\sigma_{br}^{lab} = \frac{(2\pi)^4}{3} \frac{ E(p_{n}) }{p_{n}}
\int d\Omega _1d\Omega _2dE_1 \frac{p_1 p_2^2E(p_1) E(p_2)E(p_3)}
{p_2({\sf E}-E(p_1)) - E(p_2)(\mathbf{P}-\mathbf{p}_1) \cdot \hat
{\mathbf{p}}_2 } \Gamma (W,\mathbf{k},\mathbf{q})
\left| \langle \mathbf{k}, \mathbf{q}  \Vert U_0 \Vert
\varphi_d, \mathbf{q}_0 \rangle \right| ^2
\label{eq:d.10}
\end{equation}
Inserting the explicit expression of Eqs.~(\ref{eq:d.6}) and (\ref{eq:d.7}) 
we obtain 
\begin{eqnarray}
\sigma_{br}^{lab} &=& \frac{(2\pi)^4}{3} \frac{ E(q_0)E_d(q_0)
}{2p_{n}m_d} \int d\Omega _1d\Omega _2dE_1 E(q)
\sqrt{m^2+\mathbf{k}^2} \sqrt{4(m^2+\mathbf{k}^2)+\mathbf{q}^2}
\nonumber \\ & & \times \frac{p_1p_2^2} {p_2({\sf E}-E(p_1)) -
E(p_2)(\mathbf{P}-\mathbf{p}_1) \cdot \hat {\mathbf{p}}_2 } \Gamma
(W,\mathbf{k},\mathbf{q}) \left| \langle \mathbf{k}, \mathbf{q} \Vert
U_0 \Vert \varphi_d, \mathbf{q}_0 \rangle \right| ^2,
\label{eq:d.11}
\end{eqnarray}
This gives the total invariant cross section as a five dimensional 
integral.   We have expressed it as a function of the incident 
laboratory momenta.

It follows that 
the five-fold differential cross section for exclusive breakup scattering
\begin{eqnarray}
\frac {d^5\sigma_{br}^{lab}}{ d\Omega _1d\Omega _2dE_1} 
&=& \frac{(2\pi)^4}{3} \frac{E(q_0) E_d(q_0)}{2p_{n}m_d} 
\; E(q) \sqrt{m^2+\mathbf{k}^2} 
\sqrt{4(m^2+\mathbf{k}^2)+\mathbf{q}^2}  \nonumber \\
& & \times \frac{p_1p_2^2}
{p_2({\sf E}-E(p_1)) - E(p_2)
(\mathbf{P}-\mathbf{p}_1) \cdot \hat {\mathbf{p}}_2 } 
\Gamma (W,\mathbf{k},\mathbf{q})
\left| \langle \mathbf{k}, \mathbf{q}  \Vert U_0 \Vert
\varphi_d, \mathbf{q}_0 \rangle  \right| ^2 .
\label{eq:d.12}
\end{eqnarray}

In inclusive breakup scattering only one of the outgoing particles is
detected. Thus the cross section still contains an integration over
the coordinates of the undetected particle.  In order to calculate
this cross section, it is convenient to start again from
Eq.~(\ref{eq:d.8}).  However, since we need to integrate over the
coordinates of one of the particles, we pick without loss of
generality particle $1$ as spectator and use as coordinates
\begin{eqnarray}
\sigma_{br}^{lab} &=& \frac{1}{3} \frac{ (2\pi)^4 }{v_{bt}} \int
d\mathbf{p}_1d\mathbf{k}d\mathbf{p}_{23} \left| \frac{
\partial(\mathbf{p}_2, \mathbf{p}_3) }{ \partial(\mathbf{k},
\mathbf{p}_2 + \mathbf{p}_3) }\right| \delta({\sf E}_f-{\sf E}_i)
\delta^3(\mathbf{P}_f-\mathbf{P}_i) \Gamma (W,\mathbf{k},\mathbf{q})
\left|  \langle \mathbf{k}, \mathbf{q} \Vert U_0
\Vert \varphi_d ,\mathbf{q}_0 \rangle  \right| ^2
\nonumber \\ &=& \frac{1}{3} \frac{ (2\pi)^4 }{v_{bt}} \int
d\mathbf{p}_1d\mathbf{k} \delta({\sf E}_f-{\sf E}_i) \Gamma'
(W,\mathbf{k},\mathbf{q}) | \langle \mathbf{k}, \mathbf{q} \Vert U_0
\Vert \varphi_d, \mathbf{q}_0 \rangle |^2,
\label{eq:e.1}
\end{eqnarray}
where we define
\begin{eqnarray}
\Gamma' (W,\mathbf{k},\mathbf{q}) = 
\left | \frac{ \partial(\mathbf{p}_2, \mathbf{p}_3) }{ \partial(\mathbf{k}, \mathbf{p}_2 + \mathbf{p}_3) }\right |   
\Gamma (W,\mathbf{k},\mathbf{q}).
\label{eq:e.2}
\end{eqnarray}

\noindent
Since $\delta({\sf E}_f-{\sf E}_i) = \frac{{\sf E}}{W} \delta({\sf
W}_f-{\sf W}_i) $ and $d {\sf W} / dk = 4k /
\sqrt{4(m^2+\mathbf{k}^2)+\mathbf{q}^2}$, the integration over $k$ is
eliminated leading to
\begin{equation}
\sigma_{br}^{lab} = \frac{1}{3} \frac{ (2\pi)^4 }{v_{bt}} \int
d\Omega_1 dE_1 \; k_1 E(k_1) \Gamma' (W,\mathbf{k},\mathbf{q}) \int
d\Omega_k \left| \langle \mathbf{k}, \mathbf{q} \Vert U_0 \Vert
\varphi_d, \mathbf{q}_0 \rangle \right| ^2.
\label{eq:e.3}
\end{equation}

\noindent Insert Eq.~ (\ref{eq:e.2}) gives the explicit expression for
the inclusive breakup scattering cross section
\begin{equation}
\sigma_{br}^{lab} = \frac{(2\pi)^4}{3} \frac{ E(q_0)E_d(q_0)
}{4p_{n}m_d} \int d\Omega_1 dE_1 \frac{p_1 k E(q)
(4(m^2+\mathbf{k}^2)+\mathbf{q}^2) }
{\sqrt{4(m^2+\mathbf{k}^2)+(\mathbf{P} - \mathbf{p}_1)^2}} \int
d\Omega_k \left| \langle \mathbf{k}, \mathbf{q} \Vert U_0 \Vert
\varphi_d, \mathbf{q}_0 \rangle  \right| ^2,
\label{eq:e.4}
\end{equation}
and the differential cross section 
\begin{equation}
\frac {d^3\sigma_{br}^{lab}}{ d\Omega _1 dE_1} = \frac{(2\pi)^4}{3}
\frac{ E(q_0)E_d(q_0) }{4p_{n}m_d} \; \frac{p_1 p E(q)
(4(m^2+\mathbf{k}^2)+\mathbf{q}^2)}
{\sqrt{4(m^2+\mathbf{k}^2)+(\mathbf{P} - \mathbf{p}_1)^2}} \int
d\Omega_k \left| \langle \mathbf{k}, \mathbf{q} \Vert U_0 \Vert
\varphi_d, \mathbf{q}_0 \rangle \right| ^2.
\label{eq:e.5}
\end{equation}


\clearpage

\begin{table}
\begin{tabular}{|c|r|r|r|r|r|r|r|} \hline \hline
E$_{lab}$ [GeV]     &NR [mb]  &psf  [mb]  &R-kin  [mb] & $V_0$ [mb] & $V_1$
[mb] & $V_2$ [mb] & R [mb] \\
\hline \hline
 0.01   &100027.1  &100766.0 &100605.6 &100363.2 &99288.5 &99394.2 &99276.9 \\
\hline
 0.1    &398.5     &445.0    &443.9    &418.9    &397.5   &400.1   &399.2 \\
\hline
 0.2    &167.5     &185.7    &184.5    &173.4    &163.2   &164.5   &164.1 \\
\hline
 0.5    &67.6     &83.2    &81.8     &73.5     &63.9     &65.4     &65.4 \\
\hline
 0.8    &42.9    &58.6    &57.1      &48.3     &38.7     &40.4     &40.8 \\
\hline
 1.0    &34.2    &49.7    &48.1      &39.6     &29.7     &31.5     &32.3 \\
\hline \hline
\end{tabular}
\caption{The total c.m. cross section $\sigma$ for elastic scattering calculated
from a Malfliet-Tjon type
potential. The nonrelativistic total cross section  is given in the 2nd column,
labeled NR,
the  relativistic one is given in the last column, labeled R. The
other columns give the
total cross section for elastic scattering when different relativistic features are
successively
implemented:
psf shows the effect of the relativistic phase space factor, R-kin adds the
relativistic kinematic
effects resulting from the permutation operator, for $V_0$ the
relativistic two-body LS equation
is solved with a c.m. interaction, and $V_i$ (i=1,2) denote the
approximations of the embedded interaction given
in Eqs.~(\ref{eq:v1}) and (\ref{eq:v2}).
}
\label{table-1}
\end{table}

\clearpage

\noindent

\end{document}